\documentclass[journal]{IEEEtran}

\usepackage{cite}
\usepackage{amsmath,amsthm,amssymb}
\usepackage[linesnumbered,ruled,commentsnumbered]{algorithm2e}
\usepackage{graphicx}\graphicspath{{figure/}{figure_case_study/}}
\usepackage{siunitx}
\usepackage{cases}
\usepackage[thinlines,thiklines]{easybmat}
\usepackage{physics}
\usepackage{mathtools}
\usepackage{multirow}
\usepackage[caption=false, font=footnotesize]{subfig}
\usepackage{soul}
\usepackage{color}
\usepackage[table,xcdraw]{xcolor}
\usepackage{changes}
\definecolor{blueshade}{rgb}{1,0.8,0.8}
\sethlcolor{blueshade}

\newcommand{\bI}{\mathbf{I}}
\newcommand{\bJ}{\mathbf{J}}

\newcommand{\bM}{\mathbf{M}}

\newcommand{\bp}{\mathbf{p}}

\newcommand{\bq}{\mathbf{q}}
\newcommand{\bx}{\mathbf{x}}
\newcommand{\by}{\mathbf{y}}
\newcommand{\bz}{\mathbf{z}}
\newcommand{\bff}{\mathbf{f}}
\newcommand{\bg}{\mathbf{g}}
\newcommand{\bk}{\mathbf{k}}
\newcommand{\bl}{\mathbf{l}}
\newcommand{\bu}{\mathbf{u}}
\newcommand{\bv}{\mathbf{v}}
\newcommand{\bw}{\mathbf{w}}
\newcommand{\bphi}{\mathbf{\phi}}

\newcommand{\blam}{\boldsymbol{\Lambda}}
\newcommand{\tou}{\text{out}}
\newcommand{\ti}{\text{in}}
\newcommand{\tn}{\text{node}}

\usepackage[hidelinks]{hyperref}

\begin{document}
\title{Quantify Gas-to-Power Fault Propagation Speed: A~Semi-Implicit~Simulation~Approach}
\author{Ruizhi~Yu, Suhan Zhang, Wei~Gu, Shuai Lu, Yong Sun, Baoju Li, Chengliang Hao
    \thanks{This work is supported by the National Key Research and Development Program of China (Grant No.2022YFB2404001), the National Science Fund for Distinguished Young Scholars (52325703), and the 2023 Graduate Research Innovation Plan of Jiangsu Province (Grant No. KYCX23\_0247).}
}
\maketitle
\begin{abstract}
Relying heavily on the secure supply of natural gas, the modern clean electric power systems are prone to the gas disturbances induced by the inherent rupture and leakage faults.
For the first time, this paper studies the cross-system propagation speed of these faults using a simulation-based approach.
Firstly, we establish the differential algebraic equation models of the rupture and leakage faults respectively. The boundary conditions at the fault locations are derived using the method of characteristics.
Secondly, we propose utilizing a semi-implicit approach to perform post-fault simulations. The approach, based on the stiffly-accurate Rosenbrock scheme, possesses the implicit numerical stability and explicit computation burdens. Therefore, the high-dimensional and multi-time-scale stiff models can be solved in an efficient and robust way.
Thirdly, to accurately locate the simulation events, which can not be predicted a priori, we propose a critical-time-location strategy based on the continuous Runge-Kutta approach.
In case studies, we verified the accuracy and the efficiency superiority of the proposed simulation approach. 
The impacts of gas faults on gas and power dynamics were investigated by simulation, where the critical events were identified accurately.
We found that the fault propagation speed mainly depends on the fault position and is influenced by the pipe frictions. 
The bi-directional coupling between gas and power may {lead to cascading failures}. 
\end{abstract}

\section{Introduction}
\IEEEPARstart{V}{iewed} 
as an cleaner substitution of coals, natural gas is gaining worldwide prevalence in electricity generation, tightening the coupling between natural gas networks (NGSs) and electric power systems (EPSs).
For example, the gas has dominated the fossil fuel share in UK, standing at 38.4 per cent of total generation since 2022\cite{UKElectricityStats2023}.
Gas turbines (GTs) and P2Gs are fundamental facilities bridging the gas and electric power. The GTs, owing to the high ramp rate and fast starting time, are cost-effective peak-shaving resources. By converting excessive power to gas, the P2Gs are flexible tools to smooth out the fluctuations in renewable energy\cite{Zhaijiang2022}. 

While leveraging the advantages of gas-power integrity, these facilities are also more liable to the disturbances of gas states. 
For example, in 2021, the extremely chilly weather in Texas froze the gas wells and pipes, cutting down 48.6\% of the total generation capacity\cite{HouyanqiuDezhoutingdian2022}. 
Also, the more typical rupture and leakage faults in pipes can cause pressure crash, tripping the gas turbine when its inlet-pressure-change-rate reaches a threshhold\cite{zhouEquivalentModelGas2017}.
Given this, it is of significance to quantify the cross-system fault propagation speed so that the controllers, protections and dispatchers can issue precise instructions ahead of time, stopping the blockouts from escalating and reducing costs. However, few research studies have been done about the propagation of rupture and leakage faults. This paper aims to fill this gap from the simulation point of view.

Simulation can be conceptualized as the combination of 1) forming mathematical models of certain physics and 2) solving these models within specified boundaries to test metrics.
In what follows, the literature review unfolds from the fault modeling to its simulation solutions. 
In \cite{BaoCascading2020, BaoDirection2020}, Bao et al. have incorporated the $N-1$ pipe-outage faults in dynamic simulations. The outage was modeled as tripping certain pipes from the network, which overlooked the fault dynamics itself.
Leakage is common in the gas networks, which, according to the leakage diameter, can be classified into two categories: 1) the small-hole model where the leakage mass flow is constant, and 2) the big-hole model where the leakage mass flow is pressure-dependent\cite{yuhuaEvaluationGasRelease2002}.
In \cite{OuyangShuangxiang2022}, Ouyang et al. have used the small-hole model to investigate the bi-directional impacts between gas and power by dynamic simulation.  
In \cite{ShenfuTwostageoptimal2024,TANG2023108587}, the big-hole models have been respectively used to perform the optimal dispatch and risk evaluation of the steady electricity-gas energy flow. 
In \cite{LIANG2023126898}, Liang et al. have used the big-hole model to derive the analytical leakage mass flow and pressure for dynamic risk assessment. The formulae are under the average-mass-flow-velocity (AMFV) assumption which apply only to linear pipe models.

The simulation of the EPSs is mature. Therefore the major attention has been shifted to the gas counterpart.
The partial differential equations (PDEs) governing the spatial-temporal dynamics of gas can be categorized as 1) the linear model under the AMFV assumption and 2) the original nonlinear model. Guan et al. \cite{GUAN2022118999} and Huang et al. \cite{HuangDynamicAnalysis2023} have used the finite difference methods (FDMs) to discretize the linear gas models as algebraic equations. And an analytical solution of the linear model has been derived in \cite{ZhangsuhanGas2024}. 
A comparative study of the characteristics and FDMs with respect to the linear and nonlinear models has been done in \cite{zhangsuhan2023iegs_model_sim}. It has been found that the high-order central-difference scheme produces numerical oscillation errors.
The method of characteristics has proved to be more accurate in the nonlinear case \cite{Zhaijiang2021,Zhaijiang2022,zhangsuhan2023iegs_model_sim,ZHANGsuhan2023iegs_simulation}, but the stability constraint limits the usage of big temporal step sizes, making the overall computation burden surge.
To accelerate the simulations, a sequential united method has been proposed in \cite{ZHANGsuhan2023iegs_simulation} to decrease the Jacobian dimensions. In \cite{ZHANGTONG2022,Huangxiaoli2023}, the holomorphic-embedding-based methods (HEMs) have been proposed to convert the burdensome iterations to explicit recursions. 

The unresolved issues in literature are summarized as follows.
On one hand, the leakage dynamics of the big hole case have not been studied using the nonlinear gas model. The models of rupture fault have not been established.
On the other hand, the post-fault simulation based on FDMs or characteristics is either inaccurate or time-consuming. The improvement measures cannot simultaneously eliminate iterations and ensure convergence theoretically. 
To resolve the above-mentioned problems, our work can be summarized below.
\begin{enumerate}
    \item We establish the differential algebraic equation (DAE) models of the leakage and rupture fault considering the nonlinear gas dynamics for the first time. {The WENO-3 scheme is employed to perform the high order spatial discretization of PDEs and eliminate fake oscillations\mbox{\cite{Shu1998}}.}
    \item We propose utilizing the stiffly accurate Rosenbrock method to perform the post-fault simulations. The semi-implicit method has implicit stability and explicit computation overhead, increasing the simulation efficiency and robustness.
    \item To accurately capture the simulation events, a critical-time-location strategy based on the continuous Runge-Kutta is proposed.
    \item The propagation speed of gas faults is studied with respect to the system parameters. The bi-directional impacts of the rupture and leakage on NGS and EPS dynamics are analyzed.
\end{enumerate}

The remaining part of the paper is organized as follows. Section II introduces the fault modeling. Section III presents the simulation method and the critical-time-location strategy. Section IV presents the case studies. Section V concludes.

\section{Post-Fault Simulation Modeling}
\subsection{NGS Modeling}
The governing PDEs describing the spatial-temporal transmission of pressure $\bp$ and mass flow $\bq$ are respectively \cite{ZHANGsuhan2023iegs_simulation}
\begin{equation}
    \pdv{\bp}{t}+\frac{c^2}{S}\pdv{\bq}{x}=0,\label{peqn}
\end{equation}
and
\begin{equation}
    \pdv{\bq}{t}+S\pdv{\bp}{x}+\frac{\lambda c^2\bq|\bq|}{2DS\bp}=0,\label{qeqn}
\end{equation}
where $c$ is the sound velocity, $S$ is the pipeline cross sectional area, $D$ is the pipe diameter, $\lambda$ is friction factor. These two hyperbolic PDEs can be represented by a general formula
\begin{equation}
    \pdv{\bu}{t}+\pdv{\bff(\bu)}{x}=S(\bu),
    \label{hyperbolicpde}
\end{equation}
where $\bu$ denotes $[\bp, \bq]^T$, $\bff(\bu)$ is called the flux and $S(\bu)$ is called the source term.

The continuity of mass flow and pressure are respectively described by
\begin{equation}
    \blam^{+}\cdot\bq^{\tou}+\blam^{-}\cdot\bq^{\ti}=\bq^\tn,\label{eqn_node_mass_flow_continuity}
\end{equation}
and
\begin{equation}
    p^\tn_k=p^\tou_i=p^\ti_j,\quad i\in\mathbb{E}_k^\tou,\ j\in\mathbb{E}_k^\ti,\label{eqn_node_pressure}
\end{equation}
where $\bq^{\tou/\ti}$ is the vector of outlet/inlet mass flow; $\blam^{+/-}$ is the positive/negative part of the incidence matrix $\blam$, whose entries satisfy
\begin{equation}
    \Lambda_{ij}=
    \begin{cases}
    1& \text{if pipe}\ j\ \text{flows into node}\ i\\
    -1&\text{if pipe}\ j\ \text{flows out of node}\ i\\
    0& \text{else}
    \end{cases};
\end{equation}
$\bq^\tn$ is the vector of node injection mass flow; $\bp^{\tou/\ti}$ is the vector of outlet/inlet pressure; $\bp^\tn$ is the vector of node pressure; the subscripts $i/j/k$ denote their $i/j/k$-th component; $\mathbb{E}_k^{\tou/\ti}$ is the set of pipes flowing out of/into node $k$. 
\subsubsection{Spatial Discretization}
\begin{figure}[!h]
    \centering
    \includegraphics[width=3.5in]{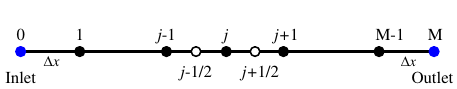}
    \caption{Discretization stencil.}
    \label{diff stencil}
\end{figure}
Denoting the spatial discretization step size by $\Delta x$, a pipe with length $L$ can be divided into $M=L/\Delta x$ sections, as shown in Fig. \ref{diff stencil}.
For each line of the vector equation \eqref{hyperbolicpde}, we discretize the spatial derivatives to obtain the ODEs
\begin{equation}
    \pdv{u_j}{t}=-\frac{1}{\Delta x}\qty(\hat{f}_{j+1/2}-\hat{f}_{j-1/2})+S(u_j), \label{weno-mol}
\end{equation}
where $j$ is the stencil index; the flux functions $\hat{f}_{i+1/2}$ and $\hat{f}_{i-1/2}$ are reconstructed by the spatially third-order weighted essentially non-oscillatory (WENO) scheme. {The scheme could suppress fake oscillations by automatically detecting the discontinuity in solutions and adaptively switching the difference stencils\mbox{\cite{Shu1998}}.}

Specifically, we first put
\[
\Bar{f}^\pm_j=\frac{1}{2}(f(u_j)\pm\alpha u_j)
\]
where $\alpha=c$ here\cite{hirsch2019numerical}. Then we can reconstruct $\hat{f}_{j+1/2}^-$ and $\hat{f}_{j-1/2}^+$ with 
\[
    \hat{f}_{j+1/2}^-=\omega_0 f_{j+1 / 2}^{(0)}+\omega_1 f_{j+1 / 2}^{(1)}
\]
and 
\[
    \hat{f}_{j-1/2}^+=\tilde{\omega}_0 f_{j-1 / 2}^{(0)}+\tilde{\omega}_1 f_{j-1 / 2}^{(1)}
\]
where
\[
    \omega_j=\frac{\alpha_j}{\alpha_0+\alpha_1}, \ \alpha_j=\frac{\gamma_j}{\left(\epsilon+\beta_j\right)^2},\ \epsilon=1e-6,\ j=0,1,
\]
\[
\beta_0=\left(\bar{f}^+_{j+1}-\bar{f}^+_{j}\right)^2,\ \beta_1=\left(\bar{f}^+_{j}-\bar{f}^+_{j-1}\right)^2,
\]
\[
    f_{j+1 / 2}^{(0)}=\frac{\Bar{f}^+_j}{2}+\frac{\Bar{f}^+_{j+1}}{2},\ 
    f_{j+1 / 2}^{(1)}=-\frac{\Bar{f}^+_{j-1}}{2}+\frac{3\Bar{f}^+_{j}}{2},
\]
\[
    \gamma_0=\frac{2}{3}, \quad \gamma_1=\frac{1}{3},
\]
\[
    \tilde{\omega}_j=\frac{\tilde{\alpha}_j}{\tilde{\alpha}_0+\tilde{\alpha}_1}, \quad \tilde{\alpha}_j=\frac{\tilde{\gamma}_j}{\left(\epsilon+\beta_j\right)^2}, \quad j=0,1,
\]
\[
    f_{j-1 / 2}^{(0)}=\frac{3 {f}^-_{j}}{2}- \frac{{f}^-_{j+1}}{2},\ f_{j-1 / 2}^{(1)}= \frac{{f}^-_{j-1}}{2} + \frac{{f}^-_{j}}{2},
\]
\[
    \tilde{\gamma}_0=\frac{1}{3},\quad\tilde{\gamma}_1=\frac{2}{3}.
\]
Moving the stencil of $\hat{f}_{j+1/2}^-$ one $\Delta x$ left and the stencil of $\hat{f}_{j-1/2}^+$ one $\Delta x$ right, we have $\hat{f}_{j-1/2}^-$ and $\hat{f}_{j+1/2}^+$. Finally, we have
\[
    \hat{f}_{j+1/2}=\hat{f}_{j+1/2}^++\hat{f}_{j+1/2}^-,\quad \hat{f}_{j-1/2}=\hat{f}_{j-1/2}^++\hat{f}_{j-1/2}^-.
\]
\subsubsection{Fault Boundary Derivation of Rupture}
\begin{figure}[!h]
    \centering
    \includegraphics[width=3.5in]{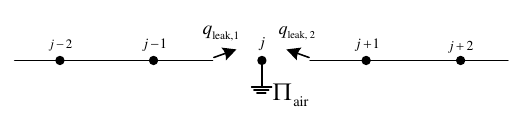}
    \caption{Diagram of rupture.}
    \label{diag_rup}
\end{figure}

As shown in Fig.\ref{diag_rup}, we denote the rupture index by $j$. When ruptures happen, $p_j$ suddenly drops to the air pressure, $p_\text{a}$. The upstream leakage mass flow, $q_\text{leak, 1}$ and the downstream leakage mass flow $q_\text{leak, 2}$ constitute the total leakage. We use the method of characteristics to approximate the leakage mass flow as follows. 

We can rewrite \eqref{peqn} and \eqref{qeqn} as
\begin{equation}
    \pdv{t}
    \begin{bmatrix}
    \bp\\
    \bq
    \end{bmatrix}
    +
    \begin{bmatrix}
    0&c^2/S\\
    S&0
    \end{bmatrix}
    \pdv{x}
    \begin{bmatrix}
    \bp\\
    \bq
    \end{bmatrix}
    =
    \begin{bmatrix}
    0\\
    -\frac{\lambda c^2\bq|\bq|}{2DS\bp}
    \end{bmatrix}.\label{eqnpq}
\end{equation}
wherein the matrix can be diagonalized as
\[
\begin{bmatrix}
0&c^2/S\\
S&0
\end{bmatrix}=
\begin{bmatrix} \frac{c}{S} & -\frac{c}{S}\\ 1 & 1 \end{bmatrix}
\begin{bmatrix} c & 0\\ 0 & -c \end{bmatrix}
\begin{bmatrix} \frac{S}{2c} & \frac{1}{2}\\ -\frac{S}{2c} & \frac{1}{2} \end{bmatrix}.
\]

We denote the diagonalization as $A=PDP^{-1}$. Then we multiply $P^{-1}$ to both sides of \eqref{eqnpq} and {neglect the source terms,} which derives
\[
    P^{-1}\pdv{t}
    \begin{bmatrix}
    \bp\\
    \bq
    \end{bmatrix}
    +
    \begin{bmatrix}
    c&0\\
    0&-c
    \end{bmatrix}P^{-1}
    \pdv{x}
    \begin{bmatrix}
    \bp\\
    \bq
    \end{bmatrix}
    =
    0.
\]

Letting 
\[\bw
=
P^{-1}    
\begin{bmatrix}
    \bp\\
    \bq
\end{bmatrix}=
\begin{bmatrix}
\frac{S}{2c} \bp+\frac{1}{2}\bq\\
-\frac{S}{2c} \bp+\frac{1}{2}\bq 
\end{bmatrix},\] 
we have
\[
    \pdv{t}
    \bw
    +
    \begin{bmatrix}
    c&0\\
    0&-c
    \end{bmatrix}
    \pdv{x}
    \bw
    =
    0,
\]
which can be written as the total derivative of $\bw$, that is
\[
    \dv{\bw}{t}=0.
\]
As a result, $\frac{S}{2c} \bp+\frac{1}{2}\bq$ and $-\frac{S}{2c} \bp+\frac{1}{2}\bq$ are all respectively invariant along the characteristics $\dv*{x}{t}=\pm c$.

We denote the invariant $S \bp+c\bq=\kappa_\text{R}(x,t)$ and $S \bp-c\bq=\kappa_\text{L}(x,t)$. Using
\[
    \kappa_\text{R}(x_j,t_1)=\kappa_\text{R}(x_{j-1},t),\quad \text{where} \ t_1=\frac{\Delta x}{c}+t
\]
and
\[
    \kappa_\text{R}(x_j,t_2)=\kappa_\text{R}(x_{j-2},t),\quad \text{where} \ t_2=\frac{2\Delta x}{c}+t,
\]
we can interpolate $\kappa_\text{R}(x_j,t)$ as 
\[
    \kappa_\text{R}(x_j,t)=2\kappa_\text{R}(x_{j-1},t)-\kappa_\text{R}(x_{j-2},t).
\]

Finally, we obtain
\begin{equation}
    q_\text{leak, 1}=2\qty(\frac{S}{c}p_{j-1}+q_{j-1})-\qty(\frac{S}{c}p_{j-2}+q_{j-2})-\frac{S}{c}p_j.\label{eqn_rup1}
\end{equation}

Using the invariant $\kappa_\text{L}(x, t)$, we have
\begin{equation}
    q_\text{leak, 2}=\qty(\frac{S}{c}p_{j+2}-q_{j+2})-2\qty(\frac{S}{c}p_{j+1}-q_{j+1})+\frac{S}{c}p_j.\label{eqn_rup2}
\end{equation}

\subsubsection{Fault Boundary Derivation of Leakage}
We denote by $d$ the diameter of the leakage hole. 
According to \cite{yuhuaEvaluationGasRelease2002}, the impact of leakage on pressure inside the pipes can be omitted and the process reaches a steady state instantaneously when $d/D< 0.2$.
In this paper, we consider the big-hole model, that is, $d/D\geq 0.2$, in which case, the leakage mass flow and pressure are time-varying.
\begin{figure}[!h]
    \centering
    \includegraphics[width=3.5in]{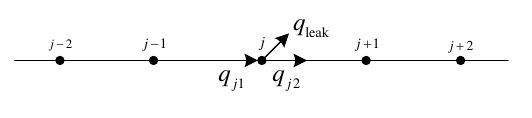}
    \caption{Diagram of leakage.}
    \label{diag_leak}
\end{figure}

Based on \cite{yuhuaEvaluationGasRelease2002}, the leakage mass flow can be calculated by, if $p_j\leq {p_\text{sw}}$, 
\begin{equation}
q_\text{leak}=0.61S_\text{h}p_j\sqrt{\frac{2M}{ZRT}\frac{k}{k-1}\left(\left(\frac{p_\text{a}}{p_j}\right)^{\frac{2}{k}}-\left(\frac{p_\text{a}}{p_j}\right)^{\frac{k+1}{k}}\right)}\label{eqn_leak1}
\end{equation} 
and, if $p_j > {p_\text{sw}}$,
\begin{equation}
q_\text{leak}=S_\text{h}p_j\sqrt{\frac{M}{ZRT}k\left(\frac{2}{k+1}\right)^{\frac{k+1}{k-1}}}\label{eqn_leak2}
\end{equation}
where $k$ is the adiabatic coefficient, $M$ is the molar mass of gas, $T$ is \SI{273.15}{\kelvin}, $Z$ is the gas compressibility factor, $R$ is the ideal gas constant, $S_\text{h}$ is the area of the leakage hole; {the switching pressure}
\[
{
p_\text{sw}=\left(\frac{2}{k+1}\right)^{-k/(k-1)}\cdot p_{a}.
} 
\]

$q_{j2}$ can be approximated by characteristics using the downstream points, that is, 
\begin{equation}
    q_{j2}=\qty(\frac{S}{c}p_{j+2}-q_{j+2})-2\qty(\frac{S}{c}p_{j+1}-q_{j+1})+\frac{S}{c}p_j.\label{eqn_leak3}
\end{equation}
Assuming that the leakage pressure $p_j$ is mainly dependent on the upstream points, we have, by characteristics,
\begin{equation}
    p_{j}=-\frac{c}{S}q_{j1}-\left(p_{j-2}+\frac{c}{S}q_{j-2}\right)+2\left(p_{j-1}+\frac{c}{S}q_{j-1}\right),\label{eqn_leak4}
\end{equation}
where $q_{j1}=q_\text{leak}+q_{j2}$.
\subsection{EPS Modeling}
We use the second-order synchronous machine model
\begin{equation}
\left\{
    \begin{aligned}
        &\dot{\omega} = \frac{P_\text{m}-P_\text{e}-D(\omega-1)}{T_j}\\
        &\dot{\delta} = (\omega-\omega_\text{COI})\omega_\text{B}\\
        &E_\text{d}'=\sin\delta\qty(U_\text{x}+r_\text{a}I_\text{x}-X_\text{q}'I_\text{y})+\cos\delta\qty(U_\text{y}+r_\text{a}I_\text{y}+X_\text{q}'I_\text{x})\\
        &E_\text{q}'=\cos\delta\qty(U_\text{x}+r_\text{a}I_\text{x}-X_\text{d}'I_\text{y})-\sin\delta\qty(U_\text{y}+r_\text{a}I_\text{y}+X_\text{d}'I_\text{x})\\
        &I_{\text{x},i}-\sum \qty(G_{ij}U_{\text{x},j}-B_{ij}U_{\text{y},j})=0\\
        &I_{\text{y},i}-\sum \qty(G_{ij}U_{\text{y},j}+B_{ij}U_{\text{x},j})=0\\
        &P_\text{e}=U_\text{x}I_\text{x}+U_\text{y}I_\text{y}+\qty(I_\text{x}^2+I_\text{y}^2)r_\text{a}
    \end{aligned}
\right.\label{epsmdl}
\end{equation}
where $\omega$ is the rotor speed; $\delta$ is the rotor angle; $P_\text{m}$ and $P_\text{e}$ are mechanical and electric power respectively; $T_j$ is the inertial constant; $D$ is the damping coefficient; $\omega_\text{COI}$ is the center-of-inertial rotor speed; $\omega_\text{B}$ is the base synchronous frequency; $E_\text{d}'$ and $E_\text{q}'$ are the d-axis and q-axis transient internal voltages; $U_\text{x}$ and $U_\text{y}$ are respectively the real and imaginary parts of voltages; $I_\text{x}$ and $I_\text{y}$ are respectively the real and imaginary parts of currents; $r_\text{a}$, $X_\text{d}'$ and $X_\text{q}'$ are generator parameters; $G$ and $B$ are respectively the conductance and susceptance matrices.
\subsection{Coupling Unit Modeling}
\subsubsection{Gas Turbine}
\begin{figure}[!h]
    \centering
    \includegraphics[width=3.5in]{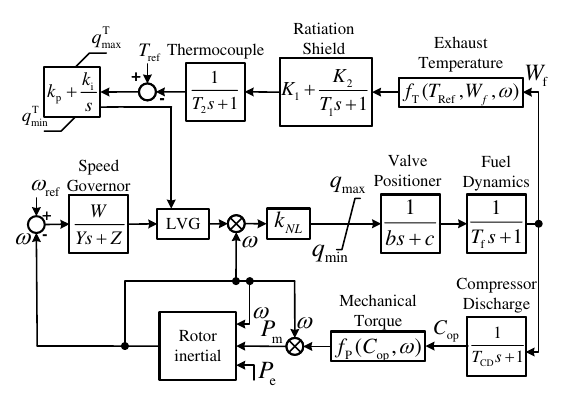}
    \caption{Diagram of gas turbine.}
    \label{diag_gt}
\end{figure}

We use the classical Rowen's gas turbine model\cite{bank2009GTmdl}, which is shown in Fig. \ref{diag_gt}. The controller adjusts the input fuel for prescribed exhaust temperature and rotor speed.
\subsubsection{P2G}
The model of P2G is \cite{fangjiakun2018}
\begin{equation}
    P_\text{P2G}=\frac{h_\text{P2G}\cdot c^2q}{\eta p},
    \label{p2g}
\end{equation}
where $P_\text{P2G}$ is the power output of P2Gs; $c$ is the sound velocity; $h_\text{P2G}$ is the enthalpy; $\eta$ is the P2G efficiency; $q$ and $p$ are respectively the mass flow generation and output pressure of P2Gs.
\section{Post-Fault Simulation Method}
The above-derived NGS fault models, the EPS models and the coupling unit models, constitute the following non-autonomous DAEs
\begin{equation}
{
\begin{bmatrix}
        \Dot{\by}(t)\\
        \mathbf{0}
    \end{bmatrix}=
    \begin{bmatrix}
        \bff(t,\by(t), \bz(t))\\
        \bg(t,\by(t), \bz(t))
    \end{bmatrix},
}\label{dae}
\end{equation}
{where $\by(t)\in\mathbb{R}^n$ and $\bz(t)\in\mathbb{R}^m$ are the vector of state and algebraic variables respectively; $n$ and $m$ are respectively the number of state and algebraic variables. The state variables in the fault-propagation scenario include 1) from NGS, the spatially discretized pressure and mass flow, and 2) from EPS, the state variables in the GT and synchronous machines. The algebraic variables include the node pressure and mass flow, the bus voltages/currents, and the algebraic variables in the GT models.
$\bff: \mathbb{R}^1\times \mathbb{R}^n\times \mathbb{R}^m\rightarrow \mathbb{R}^n$ denotes the state equations
including \mbox{\eqref{weno-mol}}, and the differential equations in both \mbox{\eqref{epsmdl}} and the GT models; $\bg: \mathbb{R}^1\times \mathbb{R}^n\times \mathbb{R}^m\rightarrow \mathbb{R}^m$ denotes the algebraic equations including \mbox{\eqref{eqn_node_mass_flow_continuity}, \eqref{eqn_node_pressure}, \eqref{eqn_rup1}-\eqref{eqn_leak4}, \eqref{p2g}}, and the algebraic equations in both \mbox{\eqref{epsmdl}} and the GT models.
}

In what follows, we present first the non-iterative semi-implicit method of the DAEs and then the critical-time-location strategy.

\subsection{Semi-implicit Solution of DAEs}
\subsubsection{Stiffly Accurate Rosenbrock Method}
Given the initial conditions $\by_0$ of $\by(t)$ and {$\bz_0$ of $\bz(t)$}, the DAE \eqref{dae} can be solved {in time interval $[t_0, t_0+h]$} with the general formulae of the $s$-stage stiffly accurate Rosenbrock method as \cite{wanner1996solving}

\begin{equation}
{
    \begin{aligned}
        &\qty(\bM-h\gamma \bJ_0)
        \begin{bmatrix}
        \bk_i\\
        \bl_i
        \end{bmatrix}\\
        =&h
        \begin{bmatrix}
        \bff(t_0+\alpha_i h,\bv_i, \bw_i)\\
        \bg(t_0+\alpha_i h,\bv_i, \bw_i)
        \end{bmatrix}
        +h\bJ_0
        \sum_{j=1}^{i-1}\gamma_{ij}
        \begin{bmatrix}
        \bk_j\\
        \bl_j
        \end{bmatrix}\\ 
        &+\gamma_i h^2 
        \begin{bmatrix}
        \pdv{\bff}{t}\\
        \pdv{\bg}{t}
        \end{bmatrix},\quad i=1,\dots, s,\\
        &
        \begin{bmatrix}
            \bv_i\\
            \bw_i
        \end{bmatrix}
        =
        \begin{bmatrix}
            \by_0\\
            \bz_0
        \end{bmatrix}
        +\sum_{j=1}^{i-1}\alpha_{ij}
        \begin{bmatrix}
        \bk_j\\
        \bl_j
        \end{bmatrix},\\
        &
        \begin{bmatrix}
            \by_1\\
            \bz_1
        \end{bmatrix}
        =
        \begin{bmatrix}
            \by_0\\
            \bz_0
        \end{bmatrix}
        +\sum_{i=1}^{s}b_i
        \begin{bmatrix}
        \bk_i\\
        \bl_i
        \end{bmatrix},\\
    \end{aligned}
    }\label{sarm1}
\end{equation}
where $s$ is the stage number; $h$ is the step size; 
\[
{
    \bJ_0=
    \begin{bmatrix}
    \pdv{\bff}{\by}&\pdv{\bff}{\bz}\\
    \pdv{\bg}{\by}&\pdv{\bg}{\bz}
    \end{bmatrix};}
\]
\begin{equation}
{
    \bM=
    \begin{bmatrix}
    \bI&\mathbf{0}\\
    \mathbf{0}&\mathbf{0}
    \end{bmatrix};}
\end{equation}
{$\bI$ is the identity matrix with the shape of $\partial \bff/\partial \by$;}
$\alpha_i=\sum\nolimits_{j=1}^{i-1}a_{ij}$, $\gamma_i=\sum\nolimits_{j=1}^{i-1}\gamma_{ij}+\gamma$, $\gamma_{ij}$, $\gamma$, $b_i$ are coefficients controlling the accuracy and numerical stability of the method. {All the derivatives are evaluated at $(t_0, \by_0, \bz_0)$.}

{The calculation is demonstrated as follows. After performing the LU factorization of $\qty(\bM-h\gamma \bJ_0)$, we obtain the factorization object $\mathbf{lu}$. Then when $i=1$, we have}
\[
{
        \begin{bmatrix}
        \bk_1\\
        \bl_1
        \end{bmatrix}\\
        =\mathbf{lu}\backslash\qty(h
        \begin{bmatrix}
        \bff(t_0,\by_0, \bz_0)\\
        \bg(t_0,\by_0, \bz_0)
        \end{bmatrix}
        +\gamma_1 h^2 
        \begin{bmatrix}
        \pdv{\bff}{t}\\
        \pdv{\bg}{t}
        \end{bmatrix}),
    }
\]
{where the $\backslash$ operator denotes the linear solver in language such as MATLAB. When $i=2$, we have}
\[
{
    \begin{aligned}
        &
        \begin{bmatrix}
        \bk_2\\
        \bl_2
        \end{bmatrix}=\mathbf{lu}\backslash\\
        &\qty(h
        \begin{bmatrix}
        \bff(t_0+\alpha_2 h,\bv_2, \bw_2)\\
        \bg(t_0+\alpha_2 h,\bv_2, \bw_2)
        \end{bmatrix}
        +h\bJ_0
        \gamma_{21}
        \begin{bmatrix}
        \bk_1\\
        \bl_1
        \end{bmatrix}+\gamma_2 h^2 
        \begin{bmatrix}
        \pdv{\bff}{t}\\
        \pdv{\bg}{t}
        \end{bmatrix}).\\
    \end{aligned}
    }
\]
{That is, only one LU factorization, which is the most time-consuming part of linear equation solutions, is required in each step.}

{Obviously, the stage-by-stage calculations of $\bk_i$ and $\bl_i$ are explicit recursions, whilst the method possesses L-stability of implicit method. As a result, the method reads \emph{semi-implicit}.} The stiff accuracy, distinguishing the method from the trivial Rosenbrock ones, ensures that both state and algebraic variables share the same order of accuracy. 

\subsubsection{Adaptive time window control}
Suppose we have $\by_1$ {and $\bz_1$} with
\[\by_1-\by(\bx_0+h)=\order{h^{q+1}}.\]
\[
{
\bz_1-\bz(\bx_0+h)=\order{h^{q+1}}.}
\]
 If we let $\hat{\by}_1=\by_0+h\sum_{i=1}^{s-1}\alpha_{si}\bk_i$ with
\[\hat{\by}_1-\by(\bx_0+h)=\order{h^{q}}.\]
{and let $\hat{\bz}_1=\bz_0+h\sum_{i=1}^{s-1}\alpha_{si}\bl_i$ with}
\[
{
\hat{\bz}_1-\bz(\bx_0+h)=\order{h^{q}}.}
\]
Then we can estimate the local error with 
\[
\begin{aligned}
    \Delta \by_1&=\order{h^q}\approx h\sum\nolimits_{i=1}^{s}\qty(b_i-\alpha_{si})\bk_i,\\
    \Delta \bz_1&=\order{h^q}\approx h\sum\nolimits_{i=1}^{s}\qty(b_i-\alpha_{si})\bl_i.
\end{aligned}
\]
To normalize the estimated error, we have
\begin{equation}
err=\qty|\frac{[\Delta \by_1\ \Delta \bz_1]^\text{T}}{Atol+Rtol\cdot |[\by_1\ \bz_1]^\text{T}|}|_\infty,
    \label{err}
\end{equation}
where $Atol$ and $Rtol$ are respectively the absolute and relative error tolerance.
Therefore, we can adjust the step size by
\begin{equation}
    h_\text{new}=h\cdot \qty(err)^{-\frac{1}{q}},\label{update_h}
\end{equation}
 Since we obtain $\hat{\by}_1$ and $\hat{\bz}_1$ without extra computation, $\hat{\by}_1$ and $\hat{\bz}_1$ are called the \emph{embedded solution} of order $q-1$.

Some of the implementations of the stiffly accurate Rosenbrock method with adaptive time window control are Rodas3\cite{SANDU19973459},  our proposed Rodas3d\cite{yu2023rodas3d}, Rodas4\cite{wanner1996solving} and Rodas5p\cite{GerdSteinebach2023}. In this paper, we utilize the famous Rodas4, which is an L-stable six-stage method with fourth order accuracy and third embedded order. Specifically, Rodas4 can be implemented with $\gamma=0.25$, 
$\alpha_{21}=0.386$, 
$\alpha_{31}=0.146$, $\alpha_{32}=0.0639$, 
$\alpha_{41}=-0.331$, $\alpha_{42}=0.711$, $\alpha_{43}=0.250$, 
$\alpha_{51}=-4.55$, $\alpha_{52}=1.71$, $\alpha_{53}=4.01$, $\alpha_{54}=-1.72$, $\alpha_{61}=2.43$, $\alpha_{62}=-3.83$, $\alpha_{63}=-1.86$, $\alpha_{64}=0.560$, $\alpha_{65}=0.250$, 
$\gamma_{21}=-0.354$, 
$\gamma_{31}=-0.134$, $\gamma_{32}=-0.0129$, 
$\gamma_{41}=1.527$, $\gamma_{42}=-0.534$, $\gamma_{43}=-1.279$, 
$\gamma_{51}=6.981$, $\gamma_{52}=-2.093$, $\gamma_{53}=-5.87$, $\gamma_{54}=0.732$, $\gamma_{61}=-2.08$, $\gamma_{62}=0.596$, $\gamma_{63}=1.702$, $\gamma_{64}=-0.089$, $\gamma_{65}=-0.379$, 
$b_1=0.348$, $b_2=0.213$, $b_3=-0.154$, $b_4=0.471$, $b_5=-0.129$, $b_6=0.25$.
\subsection{Critical Time Location}
{In this paper, we define the critical time as the moment when a significant event is \emph{triggered}, necessitating alterations to the system's structure or operational state, which in turn may pose risks to security. Efficient and accurate simulation to identify these moments can provide operators with advance warning of potential outages. Armed with this information, they are better positioned to issue timely instructions for precautionary measures or remedial actions.

Mathematically, we recognize the emergence of these events by solving}
\[
    {\bphi(t_\text{cr},\by(t_\text{cr}),\bz(t_\text{cr}))=\mathbf{0},}
\]
{where we denote by $\bphi$ the physical conditions for some events to occur, $t_\text{cr}$ the exact critical time and $\by(t_\text{cr}),\ \bz(t_\text{cr})$ the value of $\by,\ \bz$ at $t_\text{cr}$ respectively.}

We consider the cases where 1) the inlet of pressure-reducing regulators cannot satisfy the minimum requirement {$\bp_{\min}^\text{reg}$}, in which case the loads have to be cut off; 2) the mass flow supply of a gas source exceeds the maximum {$\bq_{\max}$}, so that the source type has to be converted into constant-mass-flow models{; 3) the pressure of some load crosses the boundary of normal operation interval $[\bp_{\min}^\text{load}, \bp_{\max}^\text{load}]$, impacting the gas supply quality}. To locate these events, we shall solve the equations
\begin{equation}
    p_\text{outlet}(t)-\bp_{\min}^\text{reg}=0,\label{emerg1}
\end{equation}
\begin{equation}
    {p_\text{load}(t)-\bp_{\min}^\text{load}=0,}
\end{equation}
\begin{equation}    
    {p_\text{load}(t)-\bp_{\max}^\text{load}=0,}
\end{equation}
and 
\begin{equation}
    q_\text{in}(t)-\bq_{\max}=0.\label{emerg2}
\end{equation}
in an accurate way.

Based on the continuous Runge-Kutta scheme\cite{wanner1993solving}, we can rewrite $\by_1$ and $\bz_1$ in \eqref{sarm1} as 
\begin{equation}
    \begin{bmatrix}
            \by_1\\
            \bz_1
        \end{bmatrix}
        =
        \begin{bmatrix}
            \by_0\\
            \bz_0
        \end{bmatrix}
        +\sum_{i=1}^{s}b_i(\theta)
        \begin{bmatrix}
        \bk_i\\
        \bl_i
        \end{bmatrix}
\end{equation}
where $b_i(\theta)$ is a polynomial of $\theta$. Consequently, the solution $\by_1(\theta)$ and $\bz_1(\theta)$ becomes continuous in the interval $[t_0, t_0+h]$. By substituting $\by_1(\theta)$ and $\bz_1(\theta)$ into \eqref{emerg1}-\eqref{emerg2}, we obtain the univariate polynomial equations 
\begin{equation}
    p_\text{outlet}(\theta)-\bp_{\min}=0,\label{ptheta}
\end{equation}
\begin{equation}
    {p_\text{load}(\theta)-\bp_{\min}^\text{load}=0,}
\end{equation}
\begin{equation}
    {p_\text{load}(\theta)-\bp_{\max}^\text{load}=0,}
\end{equation}
and 
\begin{equation}
    q_\text{in}(\theta)-\bq_{\max}=0,\label{qtheta}
\end{equation}
which can be solved for $\theta$, i.e., $t_\text{cr}$, easily.
Rodas4 provides a construction of continuous Runge-Kutta scheme with $\by_1(\theta)$ and $\bz_1(\theta)$ being of $\mathcal{O}(h^3)$, which is much more accurate than the simple linear interpolation.

The pseudo codes of the proposed post-fault simulation method can be found in Algorithm 1.
\begin{algorithm}
    \DontPrintSemicolon
    \SetInd{0.5em}{1em}
    \SetKwInOut{Input}{input}\SetKwInOut{Output}{output}
    \Input{$\by_0$, $\bz_0$}
    \Begin{
    Initial step size: $h\leftarrow 10^{-5}$;\;
    $t\leftarrow 0$\;
    \While{$t<T$}{
        Update $\bJ_0$ using $t$, $\by_0$ and $\bz_0$;\;
        Perform the LU factorization of $\bM-h\gamma \bJ_0$;\;
        $err\leftarrow 2$;\;
        \While{$err>1$}{
            \For{$i=1:s$}{
                Update $\bk_i$ and $\bl_i$ using \eqref{sarm1};\;
            }
            Calculate $\by_1$ and $\bz_1$ using \eqref{sarm1};\;
            Update $err$ using \eqref{err};\;
            Update $h_\text{new}$ using \eqref{update_h};\;
            $h\leftarrow h_\text{new}$;\;
        }
        \If{$\bphi(t,\by_0,\bz_0)\cdot\bphi(t+h,\by_1,\bz_1)<0$}{
        Solve \eqref{ptheta}-\eqref{qtheta} for $t_\text{cr}$\;
        break\;
        }
        $\by_0\leftarrow \by_1$;\;
        $\bz_0\leftarrow \bz_1$;\;
        $t\leftarrow t+h$;\;
    }
    }
    \caption{Main Semi-implicit Simulation Steps.}
\end{algorithm}

\section{Case Studies}
To verify the accuracy of the proposed models and simulation method, we first studied the propagation of the rupture and leakage faults in a single pipe. Then we performed the post-fault co-simulation {in coupled systems}. The cross-system propagation speed of faults and their impacts on system dynamics were investigated. {The time performance, including the time costs and the event-locating accuracy, was profiled and compared comprehensively across different methods.}

All the case studies were based on our developed simulation modeling language, Solverz 0.2.0\cite{Solverzdoc}. It provides symbolic interfaces for equation description and generates efficient numerical interfaces for the model instances. Currently, Solverz is dependent on Numpy 2.0.1, Sympy 1.12.1, Scipy 1.14.0 and Numba 0.60.0. 
We used the sparse solver, SuperLU, which is wrapped in Scipy, for the factorizations of jacobian matrices. The hardware environment is a laptop equipped with AMD Ryzen 5800H CPU and 40GB RAM.

The following schemes are used for comparison and benchmarks.

{M1: }The central difference scheme {with second order accuracy\mbox{\cite{fangjiakun2018}}}:
\[
\pdv{u}{t}=\frac{u_{i}^{j+1}+u_{i+1}^{j+1}-u_{i}^{j}-u_{i+1}^{j}}{2\Delta t}
\]
\[
\pdv{u}{x}=\frac{u_{i+1}^{j+1}+u_{i+1}^{j}-u_{i}^{j+1}-u_{i}^{j}}{2\Delta x}
\]
\[
u=\frac{u_{i}^{j+1}+u_{i+1}^{j+1}+u_{i}^{j}+u_{i+1}^{j}}{4}
\]

{M2: The Euler difference scheme with first order accuracy\mbox{\cite{zhangsuhan2023iegs_model_sim}}:}
\[
{\pdv{u}{t}=\frac{u_{i+1}^{j+1}-u_{i+1}^{j}}{\Delta t}}
\]
\[
{\pdv{u}{x}=\frac{u_{i+1}^{j+1}-u_{i}^{j+1}}{\Delta x}}
\]
\[
{u=u_{i+1}^{j+1}}
\]

{M3: The Kurganov-Tadmor scheme that is first-order spatially accurate and has the total variation decreasing property to eliminate fake oscillations\mbox{\cite{AlexanderKurganov2000}}:}
\[
    {\pdv{u_j}{t}=-\frac{1}{\Delta x}\qty(\hat{f}_{j+1/2}-\hat{f}_{j-1/2})+S(u_j),}
\]
{where $\alpha=c$,}
\[
    {\hat{f}_{j+1/2}=\frac{f(u_{j+1})-\alpha(u_{j+1}-u_j)}{2}},
\]
\[
    {\hat{f}_{j-1/2}=\frac{f(u_{j-1})-\alpha(u_{j}-u_{j-1})}{2}}.
\]
{The differential equations derived by M3 are solved by the proposed post-fault simulation method.}

{M4: the WENO-3 scheme delineated by \mbox{\eqref{weno-mol}} and solved by the proposed post-fault simulation method.}

{Benchmark: Characteristic line method}
\[
\begin{aligned}
    &p_i^{j+1}-p_{i-1}^j+\frac{c}{S}\left(q_i^{j+1}-q_{i-1}^j\right)+\frac{\lambda c^2 \Delta x}{4 D S^2} \frac{\left(q_i^{j+1}+q_{i-1}^j\right)^2}{p_i^{j+1}+p_{i-1}^j}\\
    &=0,\quad 1\leq i\leq M
\end{aligned}
\]
\[
\begin{aligned}
    &p_{i+1}^j-p_{i}^{j+1}+\frac{c}{S}\left(q_{i}^{j+1}-q_{i+1}^j\right)+\frac{\lambda c^2 \Delta x}{4 D S^2} \frac{\left(q_{i}^{j+1}+q_{i+1}^{j}\right)^2}{p_{i}^{j+1}+p_{i+1}^{j}}\\
    &=0,\quad 0\leq i\leq M-1
\end{aligned}
\]
{The characteristic method is used as the benchmark because it has no discretization error of the spatial and temporal derivatives. Also, it can only use very small step size, that is, $\Delta t=\frac{\Delta x}{c}$, where $c$ is the sound velocity around 340m/s. The method has been used as benchmark in \mbox{\cite{zhangsuhan2023iegs_model_sim,ZHANGsuhan2023iegs_simulation}}.}

\subsection{Single Pipe}
We chose a pipe with $L=51$\si{\kilo\metre}, $D=0.5901$\si{\metre} and $\lambda=0.03$. Through this pipe, a gas well with fixed pressure supplied a load with fixed mass flow. The pressure and mass flow boundary were respectively \SI{6.62}{\mega\pascal} and \SI{14}{\kilogram\per\second}. The rupture and leakage faults happened at position $x=$\SI{25.5}{\kilo\metre}. 
\subsubsection{Rupture}
\begin{figure}[!h]
    \centering
    \includegraphics[width=3.5in]{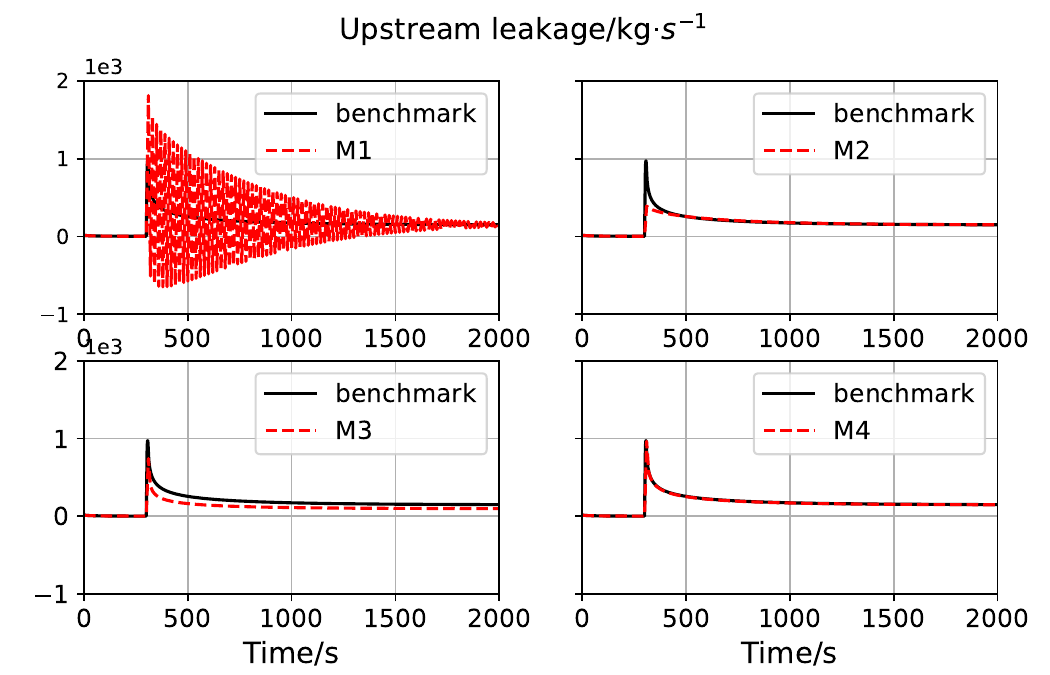}
    \includegraphics[width=3.5in]{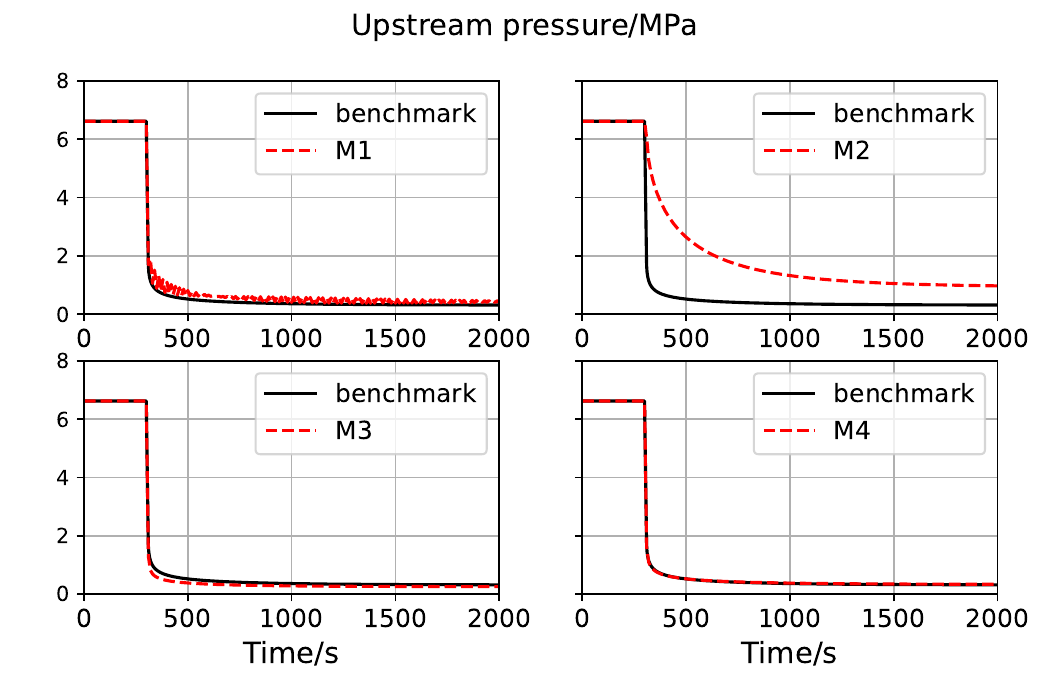}
    \includegraphics[width=3.5in]{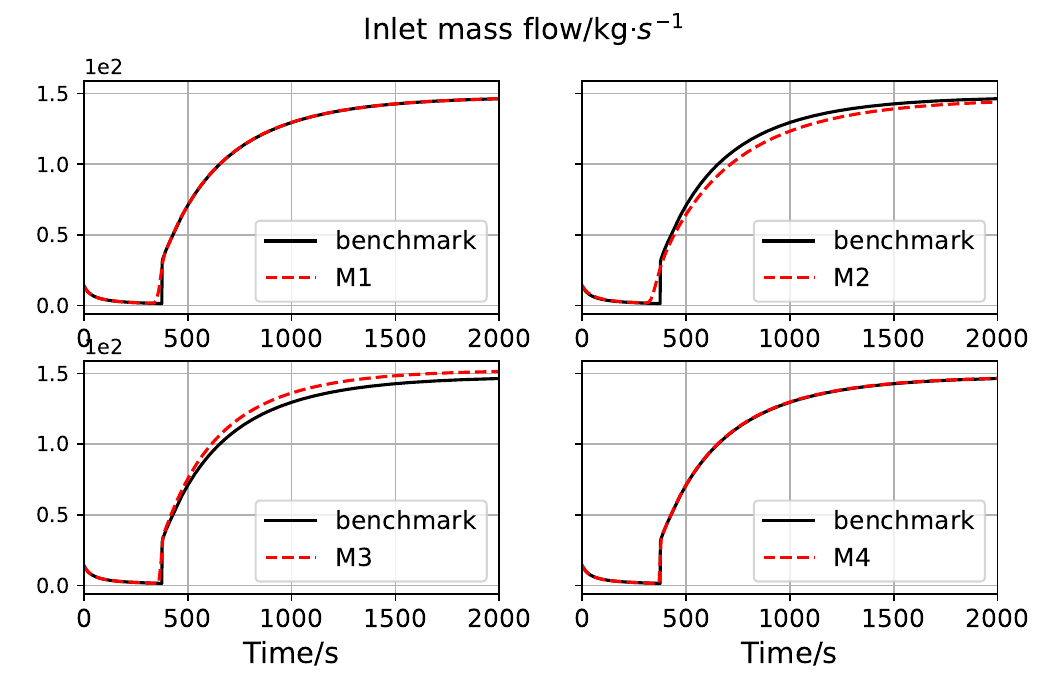}
    \includegraphics[width=3.5in]{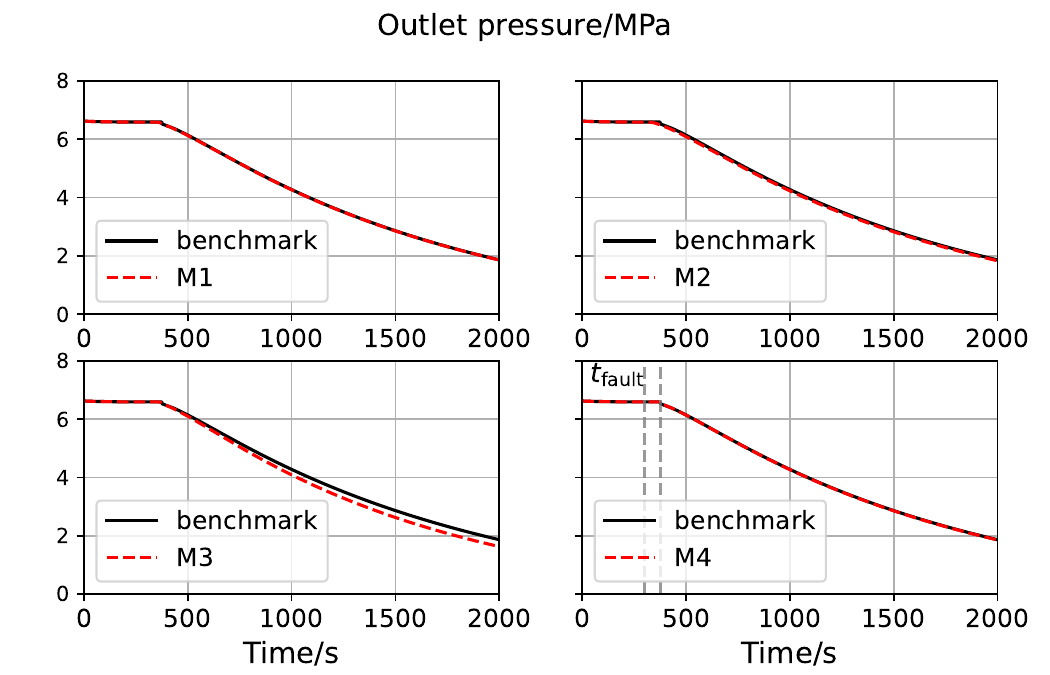}
    \caption{Pressure and mass flow variation during rupture fault.}
    \label{casei-c1}
\end{figure}

In this case, the pressure dropped from \SI{6.621}{\mega\pascal} to the standard atmospheric pressure, which is \SI{0.101}{\mega\pascal}, from $t=300$\si{\second} to $t=310$\si{\second}.
$\Delta x$ was set to be \SI{100}{\metre} for {M1 to M4}. $\Delta t$ was set to be $10$\si{\second} for {M1 to M2}. {The benchmarks were generated by characteristics with $\Delta x=50$\mbox{\si{\metre}} and $\Delta t=0.1471$\mbox{\si{\second}}.}
It is illustrated in Fig. \ref{casei-c1} that the leakage flow through the rupture spiked over $10^3$ magnitude at first and gradually decreased as the pressure level dropped.  
The outlet pressure responded to the rupture with delay around \SI{75}{\second}, which was approximately the propagation time of the sound from the rupture position. The outlet pressure level decreased to a low level within \SI{30}{\minute}.

The results by the {M1} demonstrated severe numerical oscillations with respect to the leakage mass flow and the pressure for a long duration. {M2 and M3 generated results apparently deviated from the benchmarks as shown in Fig. \mbox{\ref{casei-c1}}. This is due to the first order accuracy.} In contrast, the proposed method produced results that corresponded to the benchmarks more accurately. {The root mean square errors (RMSEs) of pivotal states are calculated in TABLE \mbox{\ref{rmse1}}. Apart from the outlet pressure on which M4 achieved slightly bigger error, M4 achieved the highest computation accuracy.}

\begin{figure}[!h]
    \centering
    \includegraphics[width=3.5in]{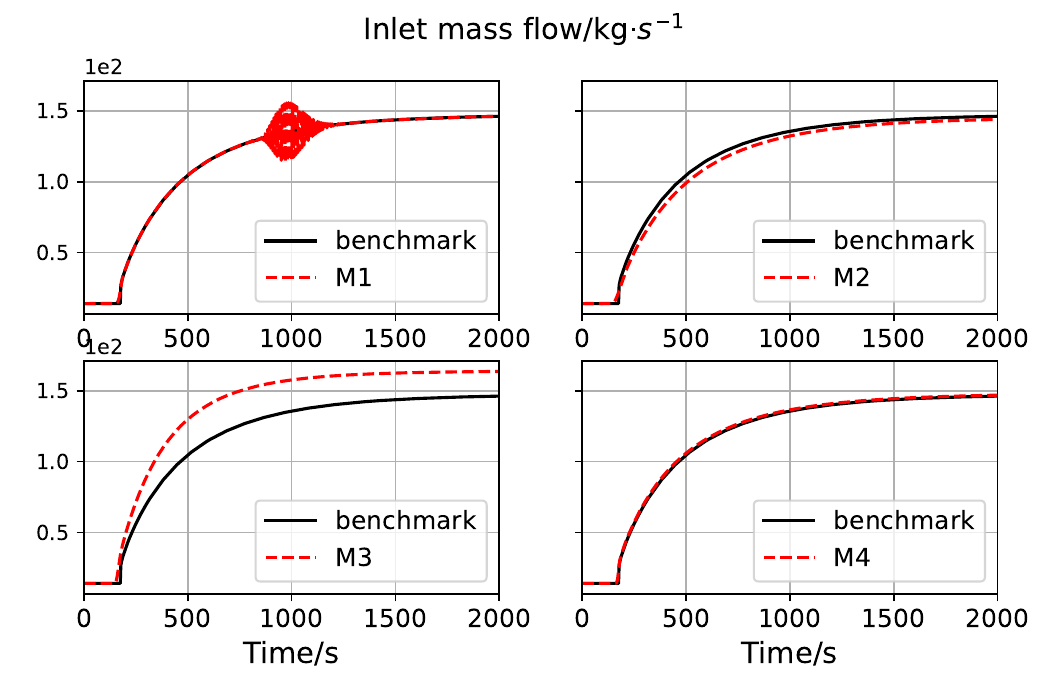}
    \includegraphics[width=3.5in]{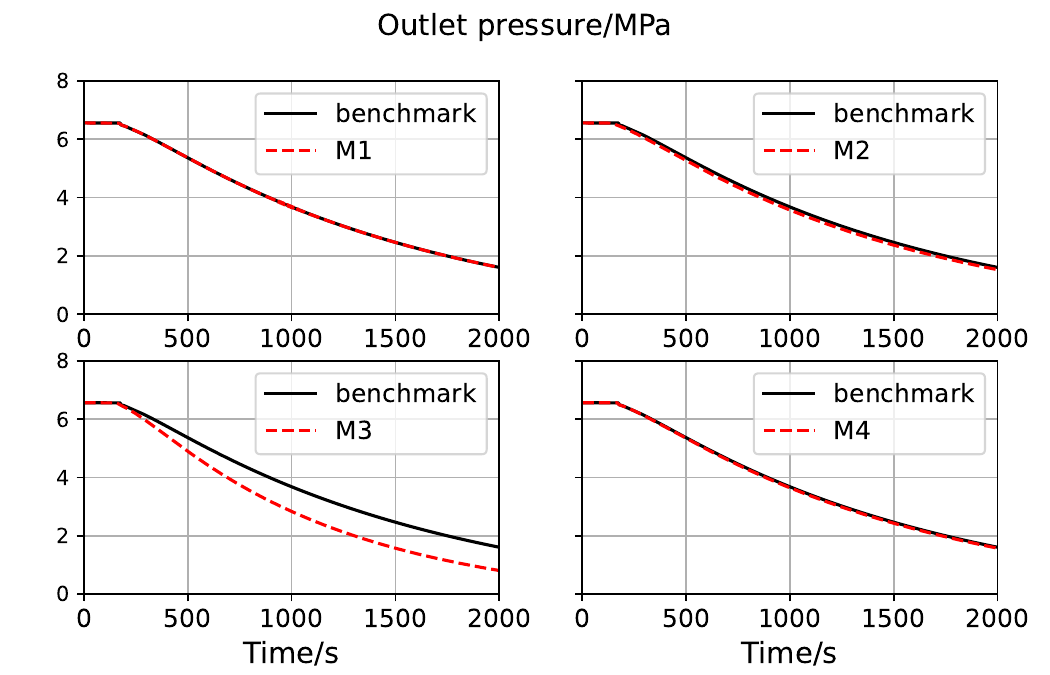}
    \caption{Pressure and mass flow variation during leakage fault.}
    \label{casei-c2}
\end{figure}

\begin{table}[]
\centering
\caption{RMSE Computed in Rupture Case}
\label{rmse1}
\begin{tabular}{lllllll}
\hline
\multicolumn{1}{c}{} & \multicolumn{1}{c}{$p_\text{out}$} & $q_\text{in}$                  & $p_\text{leak}^\text{upstream}$ & $p_\text{leak}^\text{downstream}$ & $q_\text{leak}^\text{upstream}$ & $q_\text{leak}^\text{downstream}$ \\ \hline
M1                   & \cellcolor[HTML]{C0C0C0}1.9e-3     & 1.0e0                          & 4.0e-1                          & 4.0e-1                            & 1.8e0                           & 1.9e0                             \\
M2                   & 1.1e-2                             & 1.4e0                          & 2.7e0                           & 1.2e-1                            & 9.2e-2                          & 2.7e-1                            \\
M3                   & 6.3e-2                             & 6.4e-1                         & 2.1e-1                          & 1.9e-1                            & 3.3e-1                          & 4.3e-1                            \\
M4                   & 2.5e-3                             & \cellcolor[HTML]{C0C0C0}4.8e-3 & \cellcolor[HTML]{C0C0C0}4.2e-2  & \cellcolor[HTML]{C0C0C0}6.9e-2    & \cellcolor[HTML]{C0C0C0}6.5e-2  & \cellcolor[HTML]{C0C0C0}2.6e-1    \\ \hline
\end{tabular}
\end{table}

\begin{table}[]
\centering
\caption{RMSE Computed in Leakage Case}
\label{rmse2}
\begin{tabular}{lllllll}
\hline
\multicolumn{1}{c}{} & \multicolumn{1}{c}{$p_\text{out}$} & $q_\text{in}$                  & $p_\text{leak}^\text{upstream}$ & $p_\text{leak}^\text{downstream}$ & $q_\text{leak}^\text{upstream}$ & $q_\text{leak}^\text{downstream}$ \\ \hline
M1                   & \cellcolor[HTML]{C0C0C0}2.0e-3     & 4.7e-2                         & 4.5e-1                          & 3.0e-1                            & 3.4e-2                          & 8.2e-2                            \\
M2                   & 3.1e-2                             & 5.5e-2                         & 6.7e-1                          & 2.5e-1                            & 2.4e-2                          & \cellcolor[HTML]{C0C0C0}4.6e-2    \\
M3                   & 2.8e-1                             & 1.9e-1                         & \cellcolor[HTML]{C0C0C0}1.8e-1  & 2.9e-1                            & 3.8e-1                          & 4.5e-1                            \\
M4                   & 1.2e-2                             & \cellcolor[HTML]{C0C0C0}2.9e-2 & 3.0e-1                          & \cellcolor[HTML]{C0C0C0}2.1e-1    & \cellcolor[HTML]{C0C0C0}2.3e-2  & 5.0e-2                            \\ \hline
\end{tabular}
\end{table}
\subsubsection{Leakage}
In this case, we set $d/D=0.9$ with a five-second sudden appearance of the leakage hole at $t=300$\si{\second}. 
$\Delta x$ was set to be \SI{500}{\metre} for {M1 to M4}. $\Delta t$ was set to be $5$\si{\second} for {M1 to M2}. {The benchmarks were generated by characteristics with $\Delta x=100$\mbox{\si{\metre}} and $\Delta t=0.2942$\mbox{\si{\second}}.} 
We found that the physics of leakage compared with rupture were similar.
It should be noted in Fig. \ref{casei-c2} that the inlet mass flow 
soared by several multiples within \SI{3}{\minute}. Considerable attention should be directed towards the constant-pressure gas source since its maximum output limit can be quickly attained. The switching pressure ${p_\text{sw}}$ is found to be \SI{185.07}{\kilo\pascal} but it was not met in the simulations. This is because the pressure at the leakage position was sustained by a gas well with a much more higher pressure level.

{As depicted in Fig. \mbox{\ref{casei-c2}}, M1} generated fake oscillations again, which may easily mislead the system operators over the severity and critical-time-locations of the faults. {M2 and M3, because of the first-order accuracy, produced again the less accurate results in terms of inlet mass flow and outlet pressure.} Due to the high-order spatial WENO difference, the proposed method obtained results with higher fidelity by eliminating the fake oscillations. {The RMSEs were calculated in TABLE \mbox{\ref{rmse2}}. 
It can be found that M4 illustrated the highest accuracy in most columns.
}
\subsubsection{Factors Impacting the Critical Load-cutting Time}
Since the accuracy of the proposed method was verified, we used it to study the factors impacting the propagation speed of the pressure dropping{, which can be quantified by the critical load-cutting time}. We assumed that a GT was installed at the outlet and the its minimum inlet pressure requirement was \SI{2.8}{\mega\pascal}. {The minimum pressure was set by referencing 1) paper\mbox{\cite{bank2009GTmdl}}, where a 172-MW gas turbine requires the inlet pressure to be \mbox{\SI{2.1}{\mega\pascal}}; 2) the production parameters of the Solar Gas Turbine produced by Caterpillar\mbox{\cite{SolarGasTurbine}}, where the minimum inlet pressure increase with the growth of output power. In this case and the following co-simulation scenarios, we considered GT with more than \mbox{\SI{200}{\mega\watt}} output, so that the minimum pressure was set to be higher than \mbox{\SI{2.1}{\mega\pascal}}.}

\begin{figure}[!h]
    \centering
    \includegraphics[width=3.5in]{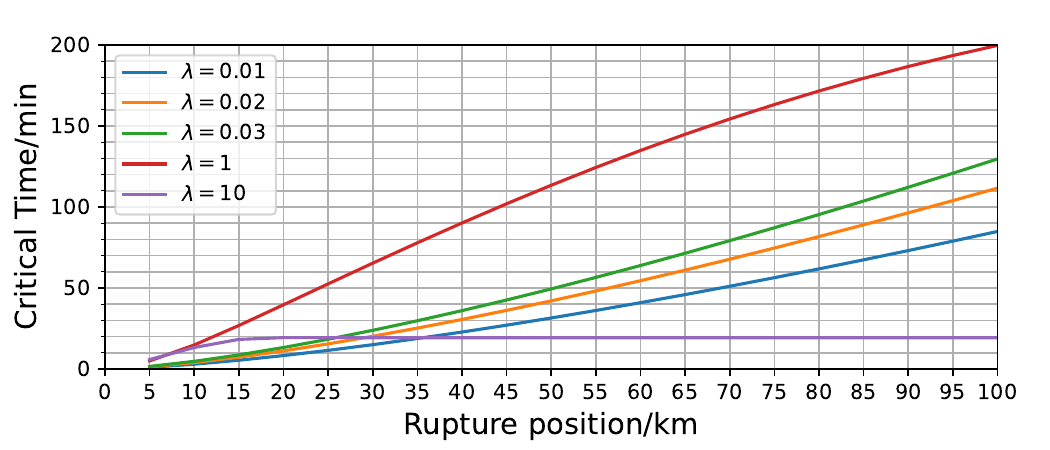}
    \includegraphics[width=3.5in]{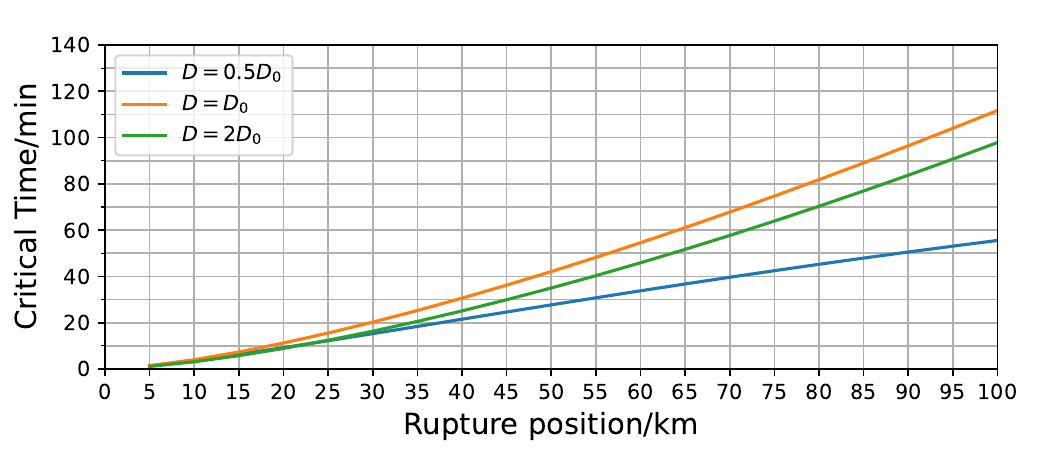}
    \caption{Critical time versus rupture position.}
    \label{casei-b1}
\end{figure}

\begin{figure}[!h]
    \centering
    \includegraphics[width=3.5in]{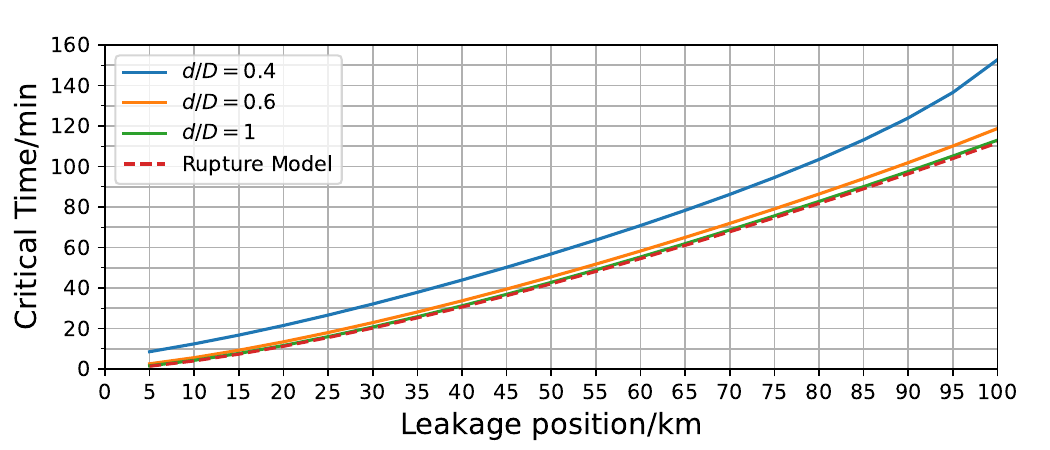}
    \caption{Critical time versus leakage position.}
    \label{casei-b2}
\end{figure}

As shown in Fig. \ref{casei-b1}-\ref{casei-b2}, the critical time increases monotonically with respect to the fault positions. During rupture faults {with $\lambda\leq 0.03$} and leakage faults with $d/D\geq 0.6$, the load has to be cut off within ten minutes if the pipe length is shorter than \SI{15}{\kilo\metre}. {The local sensitivity analysis of fault propagation speed with respect to system parameters was also performed}. {We found that when $\lambda\leq 1$} the friction coefficient $\lambda$ is negatively correlated with the critical time. {This is because there was a negative correlation between $\lambda$ and the leakage mass flow. The small $\lambda$ expedited the leakage and hence the outlet pressure decreasing. However, as indicated by the $\lambda=10$ curve in Fig. \mbox{\ref{casei-b1}}, the extremely big $\lambda$ would make the pressure level abnormally small. As a result, the critical time mainly depended on the inherent pipe frictions instead of the rupture positions. In addition, Fig. \mbox{\ref{casei-b1}} reveals that the pipe diameter $D$ made a big difference to the fault propagation speed. Pipe with $D=2D_0$ leaked more during the fault, exhausting the linpack more quickly. Pipe with $D=0.5D_0$ had big frictions, which led to more pronounced pressure drop along the pipe and hence the lower outlet pressure. For a short summary, to ascertain the $D$ and $\lambda$ that have the slowest propagation speed could help prevent the faults from being rapidly exacerbated. The proposed simulation method could act as such a tool.}

Fig. \ref{casei-b2} reveals that the leakage faults are less severe than the rupture, but it suffices to cause load-cutting as long as $d/D\geq 0.4$. The leakage case where $d/D=1$ should resemble the rupture ones and it is verified in Fig. \ref{casei-b2} by the dash line. The comparison again proves the accuracy of the proposed models. The philosophy of using characteristics to approximate fault boundaries, and the critical-time-location strategy could be used to study more complex faults.

\subsection{Post-gas-to-power-fault Co-Simulation}
\begin{figure}[]
    \centering
    \includegraphics[width=3.5in]{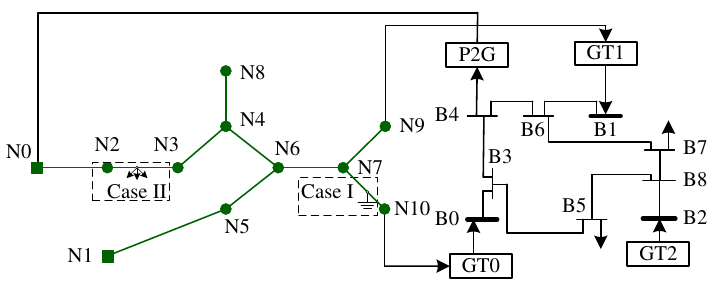}
    \caption{System diagram.}
    \label{case1-diagram}
\end{figure}

The dynamic simulations of fault propagation in the coupled NGSs and EPSs were performed using the system depicted by Fig. \ref{case1-diagram}.
GT0 and GT1 were supplied by the gas network while GT2 had local gas source. We assumed that the minimum inlet pressure requirements of them was \SI{2.5}{\mega\pascal}. Node 0 and 1 in gas networks were source node where the node 0 was supplied by the P2G in bus 4. We assumed that the maximum output of P2G was \SI{70}{\kilogram\per\second} and the node type of P2G was converted from constant pressure to constant mass-flow when the maximum was reached. The length of all pipes was \SI{51}{\kilo\metre}. Other parameters can be found in \cite{caseparameter}.
 \begin{figure}[]
    \centering
    \subfloat[]{%
    \includegraphics[width=3.5in]{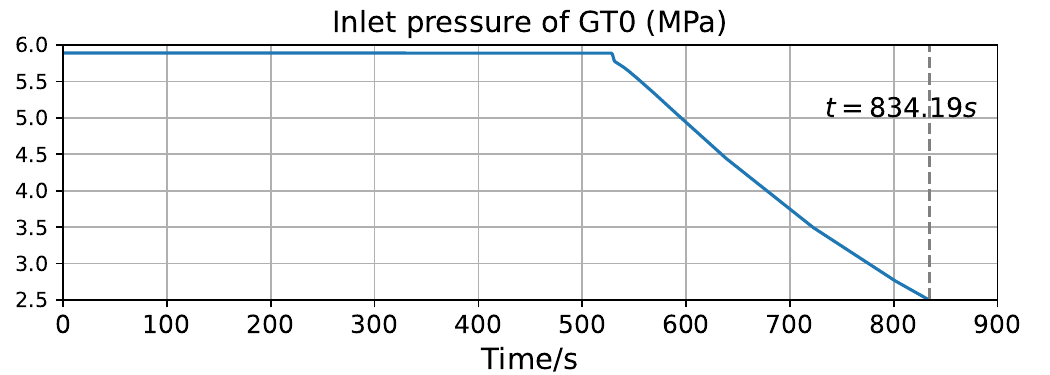}}
    \vspace{-0.3cm}
    \hfill
    \subfloat[]{%
    \includegraphics[width=3.5in]{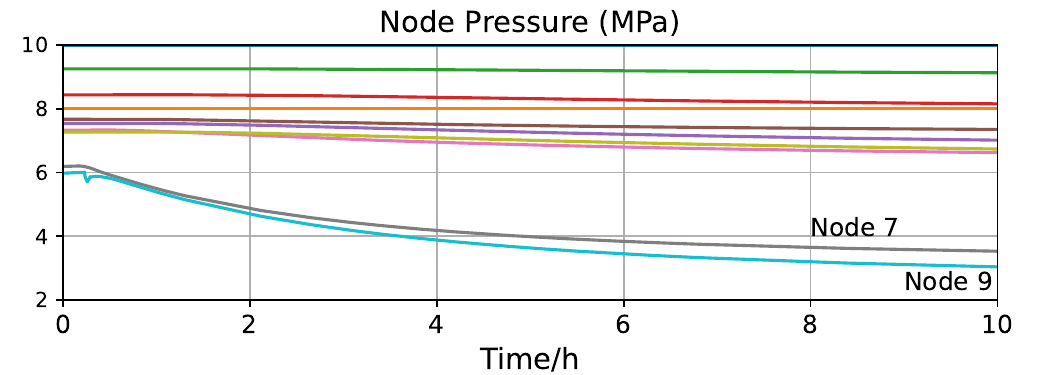}}
    \vspace{-0.3cm}
    \hfill
    \subfloat[]{%
    \includegraphics[width=3.5in]{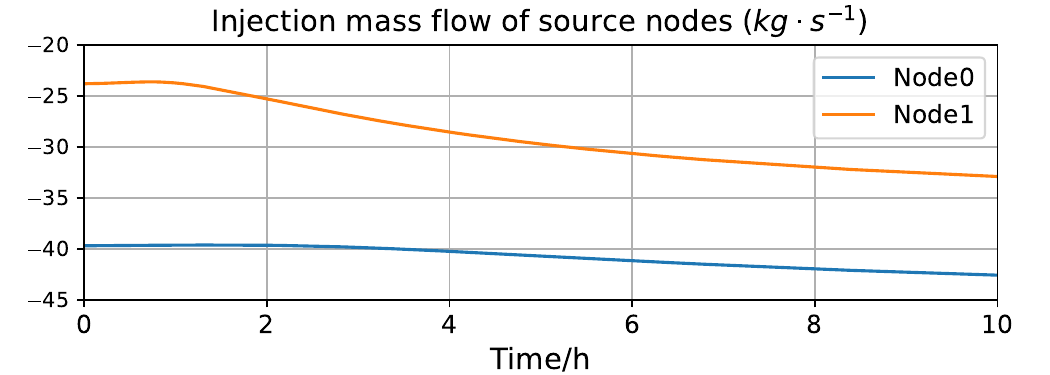}}
    \vspace{-0.3cm}
    \hfill
    \subfloat[]{%
    \includegraphics[width=3.5in]{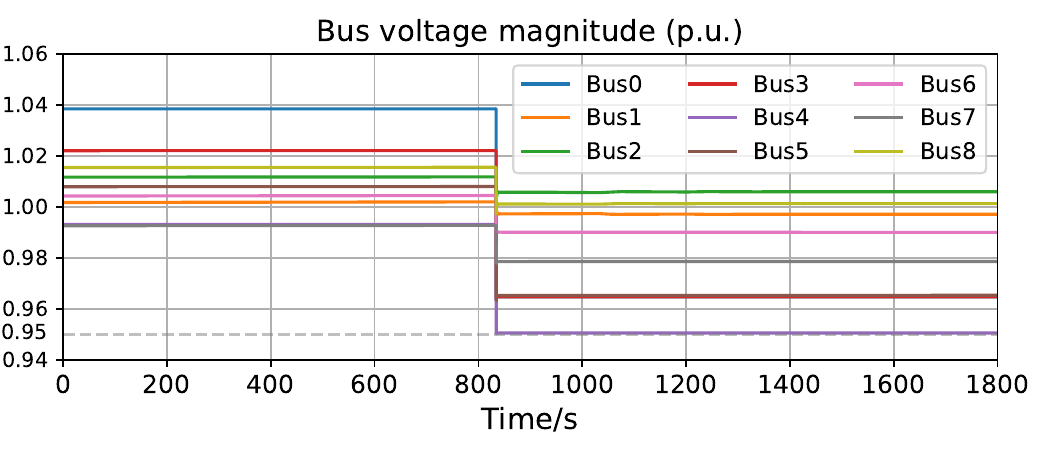}}
    \vspace{-0.3cm}
    \hfill
    \subfloat[]{%
    \includegraphics[width=3.5in]{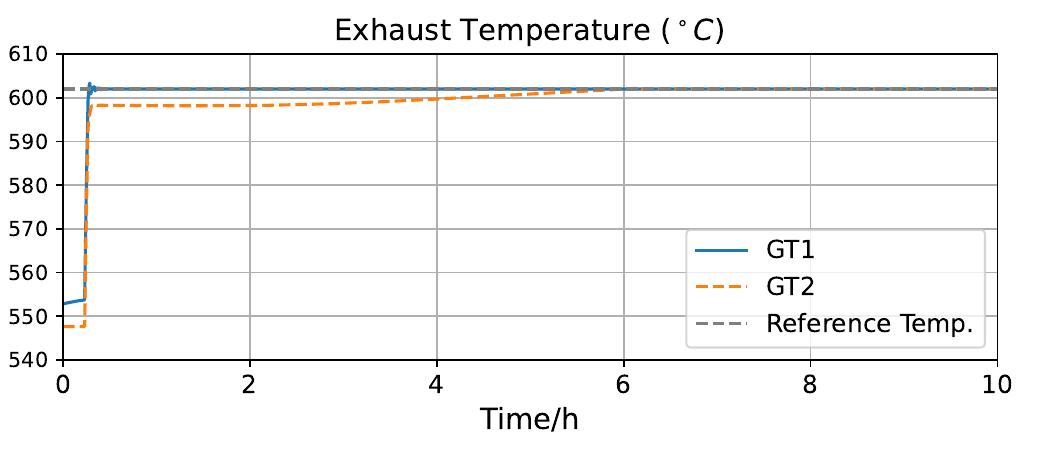}}
    \vspace{-0.3cm}
    \hfill
    \subfloat[]{%
    \includegraphics[width=3.5in]{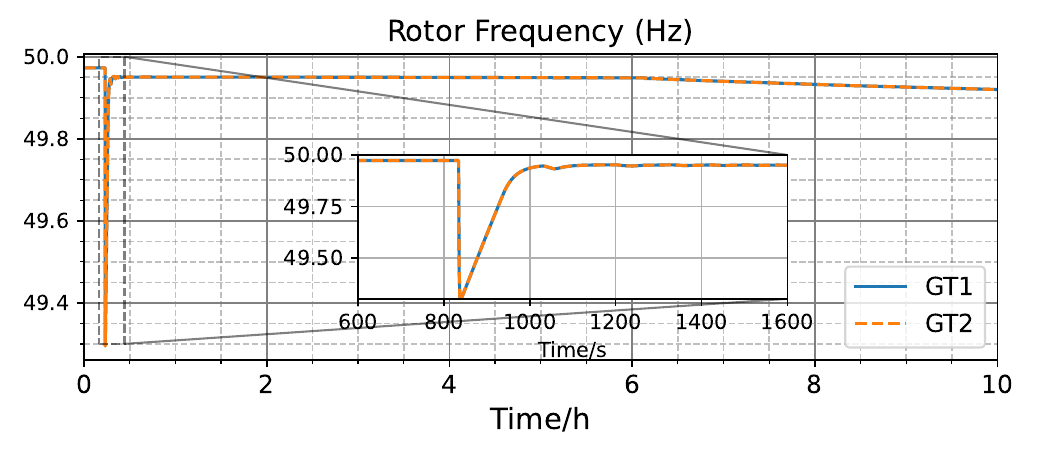}}
    \caption{Simulation results of case I.}
    \label{caseii-1}
\end{figure}
\begin{figure}[]
    \centering
    \includegraphics[width=3.5in]{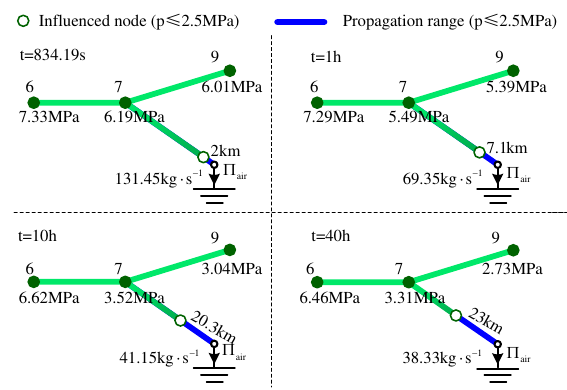}
    \caption{Fault propagation range and load influenced in Case I.}
    \label{case1-impact-load}
\end{figure}
We considered two cases where
\begin{itemize}
    \item in \emph{Case I}, the rupture happened at a distance of \SI{10.2}{\kilo\metre} from node 10 when $t=500$\si{\second};
    \item in \emph{Case II}, the leakage happened at the center between node 2 and 3 when $t=300$\si{\second}, and $d/D=0.6$.
\end{itemize}
{The simulations were performed with $\Delta x=100$m and adaptive $\Delta t$ by the proposed method. The total simulation time was set to be $10h$ and $16h$ in Case I and Case II respectively.}
\subsubsection{Case I}
The simulation results are shown in Fig. \ref{caseii-1}. By the critical-time-location strategy, it was found in Fig. \ref{caseii-1}(a) that the GT0 should be tripped at $t=834.19s$, indicating a very short interval since the rupture fault occurred. The tripping of GT0 transferred the electric power to GT1 and GT2. Hence, the bus magnitude, as well as the system frequency, suddenly dropped to a low level as shown in Fig. \ref{caseii-1}(d)(f). The voltage magnitude at bus 4 was around the lower bound, 0.95, of the EPS normal operation. It is suggested to use the AVG to increase the voltage level. The frequency was restored to the original level after the speed governor of GT1 and GT2 came into effect. Fig. \ref{caseii-1}(e) shows that the output of GT1 was restricted by the exhaust temperature limitation at the very beginning. This is because GT1 has smaller droop coefficient, so that it undertook more load variation. 

The intra-NGS fault propagation was notably slower compared to that within EPS. Fig. \ref{caseii-1}(b) reveals that while continuing to leak through the rupture position, only the pressure of node 7 and 9 came to a low level after 10 hours. From our point of view, the NGS could, due to the big linepack, run for a long time after the rupture fault. Hence, the time allotted for maintenance personnel or automatic valve devices is ample. As Fig. \ref{caseii-1}(c) shows, the major part of the leakage mass flow was undertaken by node 1 since it was closer to the rupture position. The gradually increasing P2G load, that is, the NGS source 0, was assigned to GT2 after GT1 reached its maximum output. The exhaust temperature of GT1 finally came to the reference temperature at approximately $t=6$\si{\hour}. Since then, the system frequency gradually decreased due to the mismatched generation and load.

{In this paper we focus on faults that, through pressure disturbances, impact the gas supply quality or even trip the gas turbines. The fault propagation is essentially the propagation of critical pressure that goes beyond the regulation capability of valves installed at the inlet of loads. Here, critical pressure is set to be 2.5MPa for it tripped the GTs. The fault propagation range and nodes influenced were profiled in Fig. \mbox{\ref{case1-impact-load}} using the proposed method. The critical pressure gradually propagated from the failure location towards node 7. The Fig. reveals that, following 40 hours of fault evolution, the pressure-dropping fault was confined within the pipe between node 7 and 10. This is because 1) the leakage mass flow was greatly reduced as the pressure of node 7 dropped and 2) the leakage was well sustained by the upstream source. The node 7 and 9 were working near the edge of collapse. Precautions should be taken to avoid the potential cascading failure of the GT connected to node 9.

The local sensitivity analysis of fault propagation speed with respect to pipe diameter $D$ and friction $\lambda$ was performed and profiled in TABLE \mbox{\ref{casei-sensitivity-D}} and \mbox{\ref{casei-sensitivity-lam}}. 
We found that either reducing $D$ or increasing $\lambda$ could decrease the leakage speed and hence defer the GT0's being tripped. However, the increased pressure loss along the pipes increased the risks of the cascading failure of GT1.
}

\begin{table}[]
\centering
\caption{GT-tripping Time versus Pipe Diameter in Case I}
\begin{tabular}{lll}
\hline
                      & GT0-Tripping time & GT1-Tripping time \\ \hline
$D=0.4,\lambda=0.03$ & 846.21s          &  14.47h\\
$D=0.5,\lambda=0.03$  & 834.19s          & Non-Tripping \\
$D=1,\lambda=0.03$    & 762.46s          & Non-Tripping \\ \hline
\end{tabular}
\label{casei-sensitivity-D}
\end{table}

\begin{table}[]
\centering
\caption{GT-tripping Time versus $\lambda$ in Case I}
\begin{tabular}{lll}
\hline
                      & GT0-Tripping time & GT1-Tripping time \\ \hline
$D=0.5,\lambda=0.01$  & 707.04s          & Non-Tripping \\
$D=0.5,\lambda=0.03$  & 834.19s          & Non-Tripping \\
$D=0.5,\lambda=1$     & 1646.66s         & 1700.65s         \\\hline
\end{tabular}
\label{casei-sensitivity-lam}
\end{table}
\subsubsection{Case II}
\begin{figure}[t]
    \centering
    \subfloat[]{%
    \includegraphics[width=3.5in]{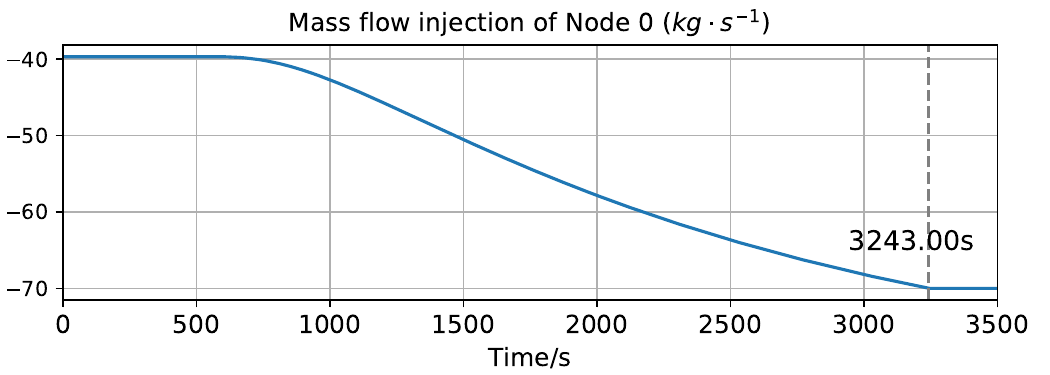}}
    \vspace{-0.3cm}
    \hfill
    \subfloat[]{%
    \includegraphics[width=3.5in]{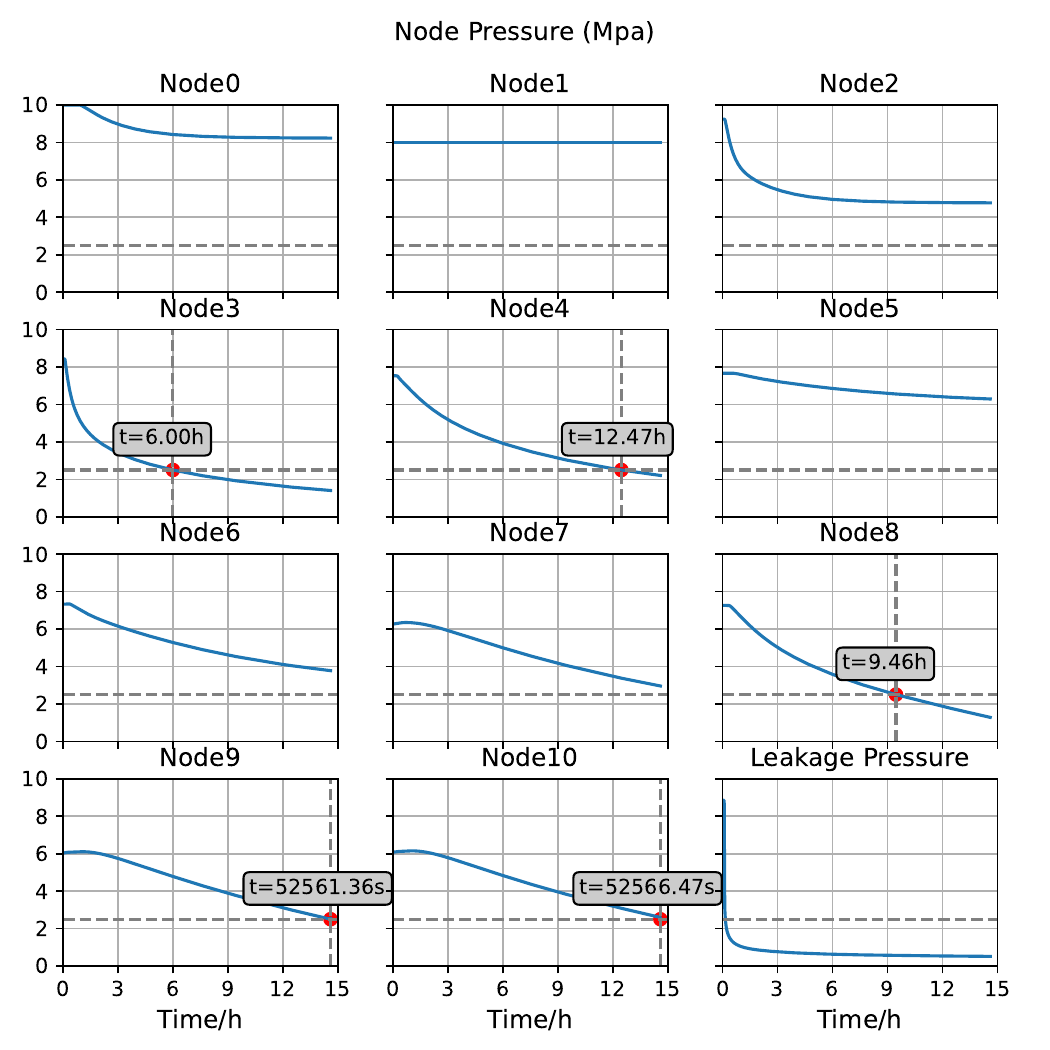}}
    \vspace{-0.3cm}
    \hfill
    \subfloat[]{%
    \includegraphics[width=3.5in]{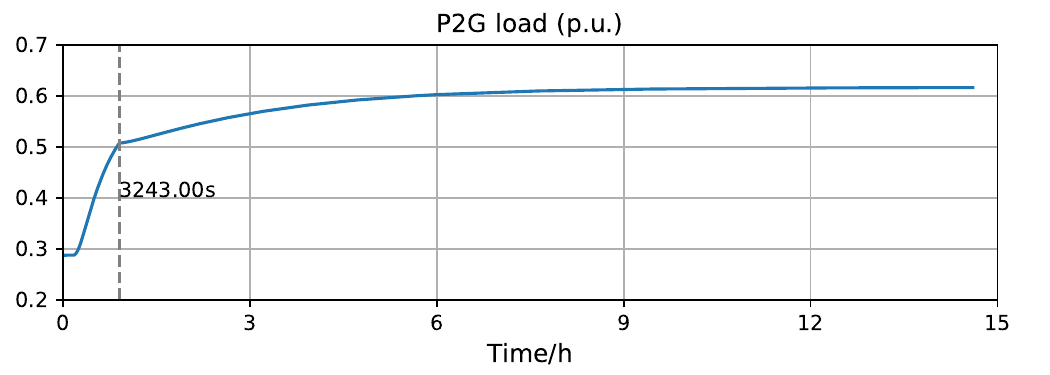}}
    \vspace{-0.3cm}
    \hfill
    \subfloat[]{%
    \includegraphics[width=3.5in]{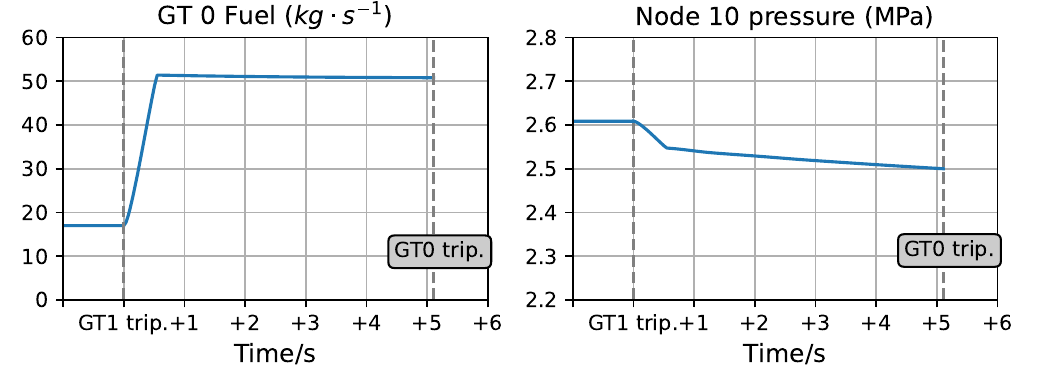}}
    \vspace{-0.3cm}
    \hfill
    \caption{Simulation results of case II.}
    \label{caseiii}
\end{figure}

The simulation results are shown in Fig. \ref{caseiii}. By the critical-time-location strategy, it was found in Fig. \ref{caseiii}(a) that the P2G reached the maximum output at $t=3243.00s$, indicating the node type conversion from constant pressure to constant mass-flow. We again used the critical-time-location strategy to find the critical time when certain node pressure was below \SI{2.5}{\mega\pascal} during the propagation of leakage fault. The critical time of these high-risk nodes are annotated in Fig. \ref{caseiii}(b). The pressure of node 3, 4 and 8 decreased to a lower level earlier than the others{, to the detriment of the gas supply quality on these nodes. This is because} they were at the downstream locations of the fault. The pressure of node 9 and 10 was also liable to the leakage fault because of its lower initial pressure and the increasing load level indicated in Fig. \ref{caseiii}(d). On the contrary, the pressure of node 5, 6 and 7 was sustained by the source node 1.
\begin{figure}[]
    \centering
    \includegraphics[width=3.5in]{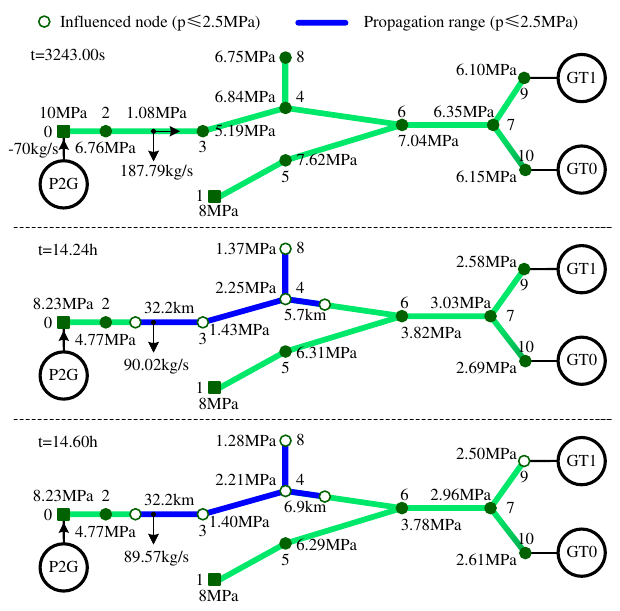}
    \caption{Fault propagation range and load influenced in Case II.}
    \label{case2-impact-load}
\end{figure}

{The fault propagation range and nodes influenced were illustrated in Fig. \mbox{\ref{case2-impact-load}}. The critical pressure gradually moved from node 4 to node 6 and from node 3 to node 2, impacting the normal operation of the downstream area of the P2G source.} 

{We found by simulation that the bi-directional of gas and electricity led to cascading failures of EPSs. As revealed by Fig. \mbox{\ref{case2-impact-load}}, it took \mbox{\SI{0.36}{\hour}} for node 9 to decrease from \mbox{\SI{2.58}{\mega\pascal}} to \mbox{\SI{2.50}{\mega\pascal}}, causing the tripping of GT1. However, we found in Fig. \mbox{\ref{caseiii}} (b)(d) that GT0, the inlet pressure of which was \mbox{\SI{2.61}{\mega\pascal}} at the GT1 failure time, was tripped only \mbox{\SI{5.1}{\second}} later.}
{On one hand, the P2G load in EPS, as depicted in Fig. \mbox{\ref{caseiii}}(c), was continually increasing. It can be explained by \mbox{\eqref{p2g}} that the load was first subject to the increasing mass flow injection but then to the decreasing pressure after the P2G node type conversion. }
{On the other hand, the tripping of GT1 made the speed governor shift the increasing EPS load to GT0 and GT2. Hence their required gas fuel surged within one second as shown in Fig. \mbox{\ref{caseiii}}(d), decreasing the inlet pressure of GT0 rapidly. Consequently, the GTs quit operation in quick succession and the EPSs finally unraveled.}

\begin{table}[]
\centering
\caption{Critical Time versus Pipe Diameter in Case II}
\begin{tabular}{lll}
\hline
                      & P2G-Saturation time & GT1-tripping time \\ \hline
$D=0.4,\lambda=0.03$ & Non-Saturation       &  Sim. Failed at $t=7.00h$\\
$D=0.5,\lambda=0.03$  & 3242.97s            & 14.60h \\
$D=0.6,\lambda=0.03$  & 1106.68s & Non-tripping \\ \hline
\end{tabular}
\label{caseii-sensitivity-D}
\end{table}

\begin{table}[]
\centering
\caption{GT-tripping Time versus $\lambda$ in Case II}
\begin{tabular}{lll}
\hline
                      & P2G-Saturation time & GT1-tripping time \\ \hline
$D=0.5,\lambda=0.01$  & 717.52s  & Non-tripping \\
$D=0.5,\lambda=0.03$  & 3242.97s         & 14.60h \\
$D=0.5,\lambda=0.1$     & Non-Saturation         & Sim. Failed at $t=6.06h$         \\\hline
\end{tabular}
\label{caseii-sensitivity-lam}
\end{table}

{In this case, the local sensitivity analysis of fault propagation speed with respect to pipe diameter $D$ and friction $\lambda$ was performed and profiled in TABLE \mbox{\ref{caseii-sensitivity-D}} and \mbox{\ref{caseii-sensitivity-lam}}. We found that when $\lambda=0.03$, $D\geq 0.5$ was big enough to cause the P2G-saturation. Also, when $D=0.5$, $\lambda\leq 0.03$ could cause the P2G-saturation. This is because either big $D$ or small $\lambda$ increased the leakage speed and hence accelerated the saturation of P2G. Similar to Case I, big $D$ and small $\lambda$ decreased the pressure loss along the pipes. From TABLE \mbox{\ref{caseii-sensitivity-D}}, we could see that either $D = 0.6$ and $\lambda=0.03$, or $D=0.5$ and $\lambda = 0.01$ would not cause the tripping of GT1. The case where $D=0.4$ and $\lambda=0.03$ and the case where $D=0.5$ and $\lambda=0.1$ should be explained here since the simulation failed because of low pressure level. We found that in both cases there was zero pressure at node 8, making our fault models inapplicable. Additionally, the pressure levels in the system were extremely low due to significant friction, and the GT-1 was nearly tripped.}
\subsubsection{Time Performance}
{We compare the simulation time costs and the accuracy of event location based on}
\begin{itemize}
    \item {Benchmark, which alternatively solves the NGS models with characteristics and $\Delta t_\text{ngs}=0.2941s$, and the EPS counterpart with the implicit trapezoidal method and $\Delta t_\text{eps}=5e-4s$. If the trapezoidal method failed, then EPS part was solved by Rodas4 instead.}
    \item {EU-TZ, which alternatively solves 1) the NGS models with the Euler difference scheme and $\Delta t_\text{ngs}=10s$, and 2) the EPS counterpart with the implicit trapezoidal method and $\Delta t_\text{eps}=1e-3s$ in Case I and $\Delta t_\text{eps}=5e-2$ in Case II.}
    \item {CD-TZ, which alternatively solves 1) the NGS models with the central difference scheme and $\Delta t_\text{ngs}=10s$, and 2) the EPS counterpart with the implicit trapezoidal method and $\Delta t_\text{eps}=1e-3s$ in Case I and $\Delta t_\text{eps}=5e-2$ in Case II.}
    \item {KT-RD, which performs the semi-discretization of PDEs with the above-mentioned first-order Kurganov-Tadmor scheme, and solves the DAEs with Rodas4 from Section III.}
    \item {WN-RD, which performs the semi-discretization of PDEs with the WENO3 scheme from Section II, and solves the DAEs with Rodas4 from Section III.}
\end{itemize}

{The spatial step sizes for all methods were set to be $\Delta x=100m$. The $Atol$ and $Rtol$ of KT-RD and WN-RD were set to be $1e-3$ and $1e-1$ respectively. }

{The event location results and the simulation time costs in cases I are profiled in Table \mbox{\ref{ELR1}} and Table \mbox{\ref{Time Costs1}} respectively. Apart from KT-RD, the methods EU-TZ, CD-TZ, and WN-RD successfully identified the tripping of GT0 with a deviation of less than five seconds. Notably, WN-RD led this group by a narrow margin of 2.81 seconds. When it came to the pre-GT-tripping simulation, WN-RD demonstrated a slight edge over EU-TZ and CD-TZ in terms of computational time. Nevertheless, EU-TZ and CD-TZ encountered robustness issues during the post-GT-tripping simulation, failing to achieve convergence within the specified error tolerance. Thanks to their non-iterative and stiffly accurate characteristics, both WN-RD and KT-RD outperformed the benchmark in simulation efficiency, effectively addressing the non-convergence problem. KT-RD, in particular, surpassed WN-RD in efficiency due to its use of a three-point stencil for spatial discretization, as opposed to the five-point stencil employed by WN-RD. This difference made the Jacobian matrix of KT-RD significantly sparser, thereby saving time on LU factorization at each step of the simulation process.}

{The event location results and the simulation time costs in cases II are profiled in Table \mbox{\ref{ELR2}} and Table \mbox{\ref{Time Costs2}} respectively. Regarding the location of the P2G saturation event, WN-RD demonstrated the highest accuracy, with deviations of 1.12 seconds. In contrast, EU-TZ and CD-TZ exhibited much larger deviations, which are more than 70 seconds. As highlighted by the single-pipe simulations, these significant errors can be attributed to the artificial oscillations inherent in the central difference scheme and the low-order accuracy of the Euler difference scheme, both of which can lead to substantial inlet mass flow errors. Furthermore, the event locations by EU-TZ and CD-TZ are constrained by the minimum resolution of the discrete temporal step sizes, introducing additional inaccuracies. Conversely, KT-RD and WN-RD provide continuous solutions by constructing and solving an equation for the critical time using interpolation. This approach, combined with the embedded solution, ensures a high level of accuracy in determining the event location. 

For the GT1-tripping event, WN-RD achieved the most precise results. Results by EU-TZ and CD-TZ exhibit minor deviations, on the order of minutes, whereas KT-RD showed a significantly larger discrepancy, with an error exceeding half an hour. In terms of computational efficiency, WN-RD was more than fifteen times faster than both EU-TZ and CD-TZ. Furthermore, WN-RD outperformed the benchmark by a factor of 100 in speed, all while delivering results that most closely approximated the benchmark's accuracy.

This performance can be attributed to three key factors: 1) Efficiency in Iterations: WN-RD does not require Newton iterations, meaning the Jacobian is factorized only once per step; 2) Higher Order Schemes: By employing higher-order spatial (WENO3) and temporal (Rodas4) schemes, WN-RD can locate the events more precisely; 3) Adaptive Time Window: WN-RD uses larger average step sizes through its adaptive time window control strategy, resulting in fewer overall computation steps.
}

\begin{table}[!h]
	\centering
        \caption{Event Location Results in Case I}
        \begin{tabular}{lc}
        \hline
        Method& GT0 Tripping \\ \hline
        Benchmark  & 837.00s   \\
        EU-TZ   & 830.00s  \\
        CD-TZ   & 840.00s   \\
        KT-RD   & 794.90s   \\
        WN-RD   & 834.19s   \\ 
        \hline
        \end{tabular}
        \label{ELR1}
\end{table}
\begin{table}[!h]
	\centering
        \caption{Time Costs in Case I}
        \begin{tabular}{lccc}
		\hline
  		 Method&Pre-GT-Tripping&Post-GT-Tripping& Total \\ 
   \hline
        Benchmark&322.39s&5604.19s&5926.58s\\
        EU-TZ&12.39s&Non-converge&N\textbackslash A\\
        CD-TZ&14.33s&Non-converge&N\textbackslash A\\
        KT-RD&2.01s&12.70s&14.71s\\
	WN-RD&9.62s&83.66s&93.28s\\ 
        \hline
	\end{tabular}
	\label{Time Costs1}
\end{table}
\begin{table}[!h]
	\centering
        \caption{Event Location Results in Case II}
        \begin{tabular}{lcc}
        \hline
        Method& P2G Saturation&GT1 Tripping \\ \hline
        Benchmark  & 3244.12s &14.60h  \\
        EU-TZ   & 3350.00s &14.68h\\
        CD-TZ  &  3320.00s &14.69h\\
        KT-RD   & 3037.06s &13.91h    \\
        WN-RD & 3243.00s &14.60h  \\
        \hline
        \end{tabular}
        \label{ELR2}
\end{table}
\begin{table}[!h]
	\centering
        \caption{Time Costs in Case II}
        \begin{tabular}{lccc}
		\hline
        Method&Pre-Saturation&Post-Saturation& Total \\
        \hline
        Benchmark&173.62s&9639.87s&9813.49s\\
        EU-TZ&117.10s&1269.26s&1386.36s\\
        CD-TZ&120.78s&1340.81s&1461.59s\\
        KT-RD&15.65s&1.65s&17.3s\\
	WN-RD&88.95s&1.05s&90.00s\\ 
 \hline
	\end{tabular}
	\label{Time Costs2}
\end{table}

\subsection{Post-power-to-gas-fault Co-simulation}
{In this case, we set the three-phase short-circuit on bus 6 at $t=1.002$\mbox{\si{\second}} and cleared it \mbox{\SI{30}{\milli\second}} later.}

\begin{figure}[t]
    \centering
    \subfloat[]{%
    \includegraphics[width=3.5in]{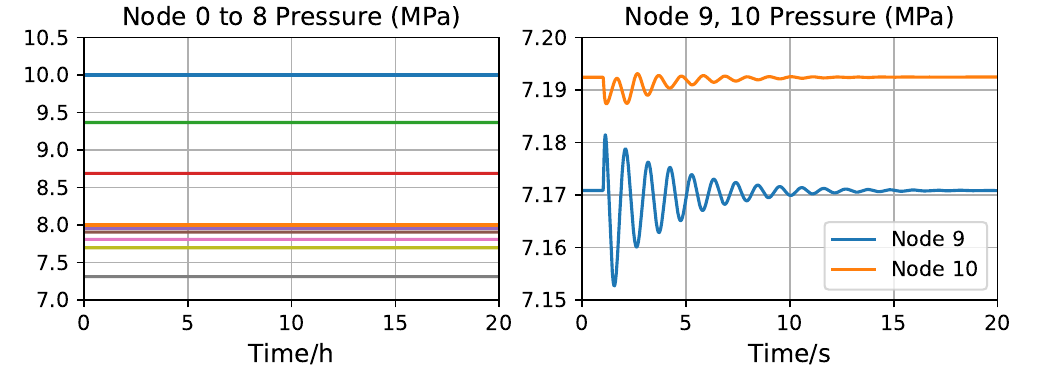}}
    \vspace{-0.3cm}
    \hfill
    \subfloat[]{%
    \includegraphics[width=3.5in]{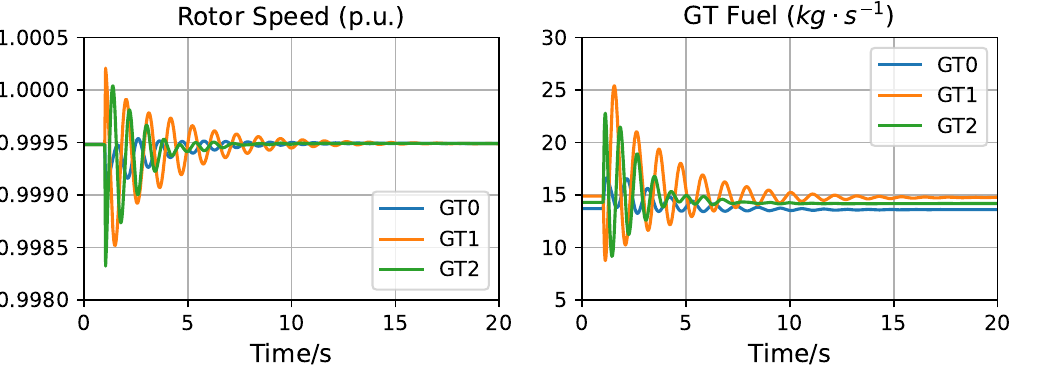}}
    \vspace{-0.3cm}
    \hfill
    \subfloat[]{%
    \includegraphics[width=3.5in]{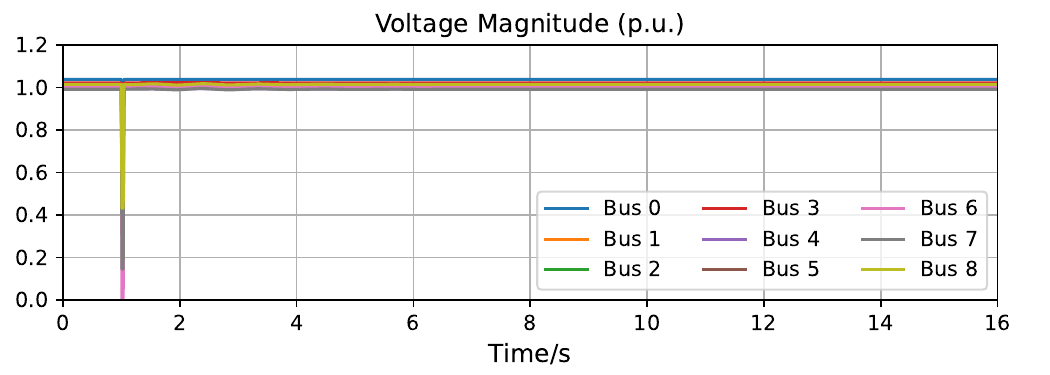}}
    \vspace{-0.3cm}
    \hfill
    \caption{Results of post-power-to-gas-fault co-simulation.}
    \label{caseiv}
\end{figure}

{The simulation results are shown in Fig. \mbox{\ref{caseiv}}. In EPS side, the fault led to voltage plunging and GT rotor's oscillation. But, after clearing the fault, the disturbed states were soon reinstated within \mbox{\SI{20}{\second}}. The GT fuel was impacted by the electrical power during the fault, which caused the inlet pressure's ripples. The ripples on node 9 and 10 are trivial compared with those in leakage and rupture cases, which were filtered by pipe frictions and contributed to no pressure disturbances on other nodes as shown in Fig. \mbox{\ref{caseiv}}(a). As a result, the three-phase short circuit fault has no significant influences on the NGS states, because the protections can clear the fault rapidly and the EPS dynamics are fleeting.}
\begin{figure}[t]
    \centering
    \subfloat[]{%
    \includegraphics[width=3.5in]{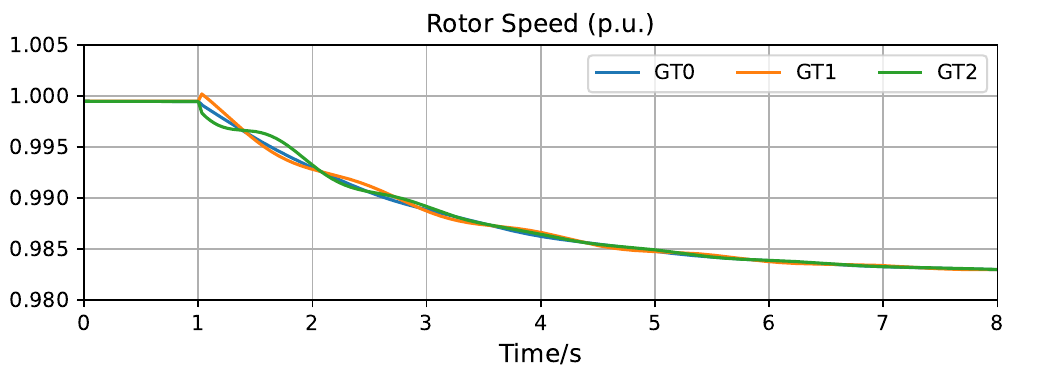}}
    \vspace{-0.3cm}
    \hfill
    \subfloat[]{%
    \includegraphics[width=3.5in]{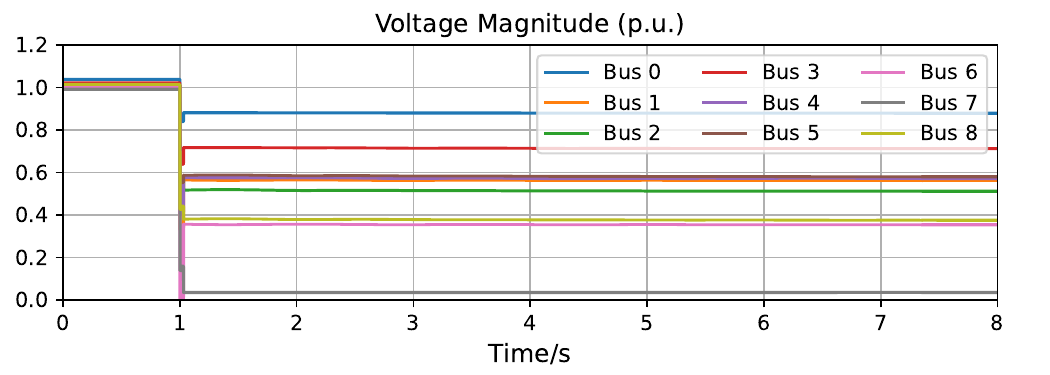}}
    \vspace{-0.3cm}
    \hfill
    \caption{Low voltage solutions by fixed-step implicit trapezoidal method.}
    \label{caseiv2}
\end{figure}

{We used the implicit trapezoidal method with fixed step size $1e-3$ to perform the simulation. The results in Fig. \mbox{\ref{caseiv2}} indicate that the trapezoidal method converged to the low voltage solutions, which again attest to the numerical robustness of the proposed simulation method.}

\subsection{Large system}
\begin{figure}[]
    \centering
    \includegraphics[width=3.5in]{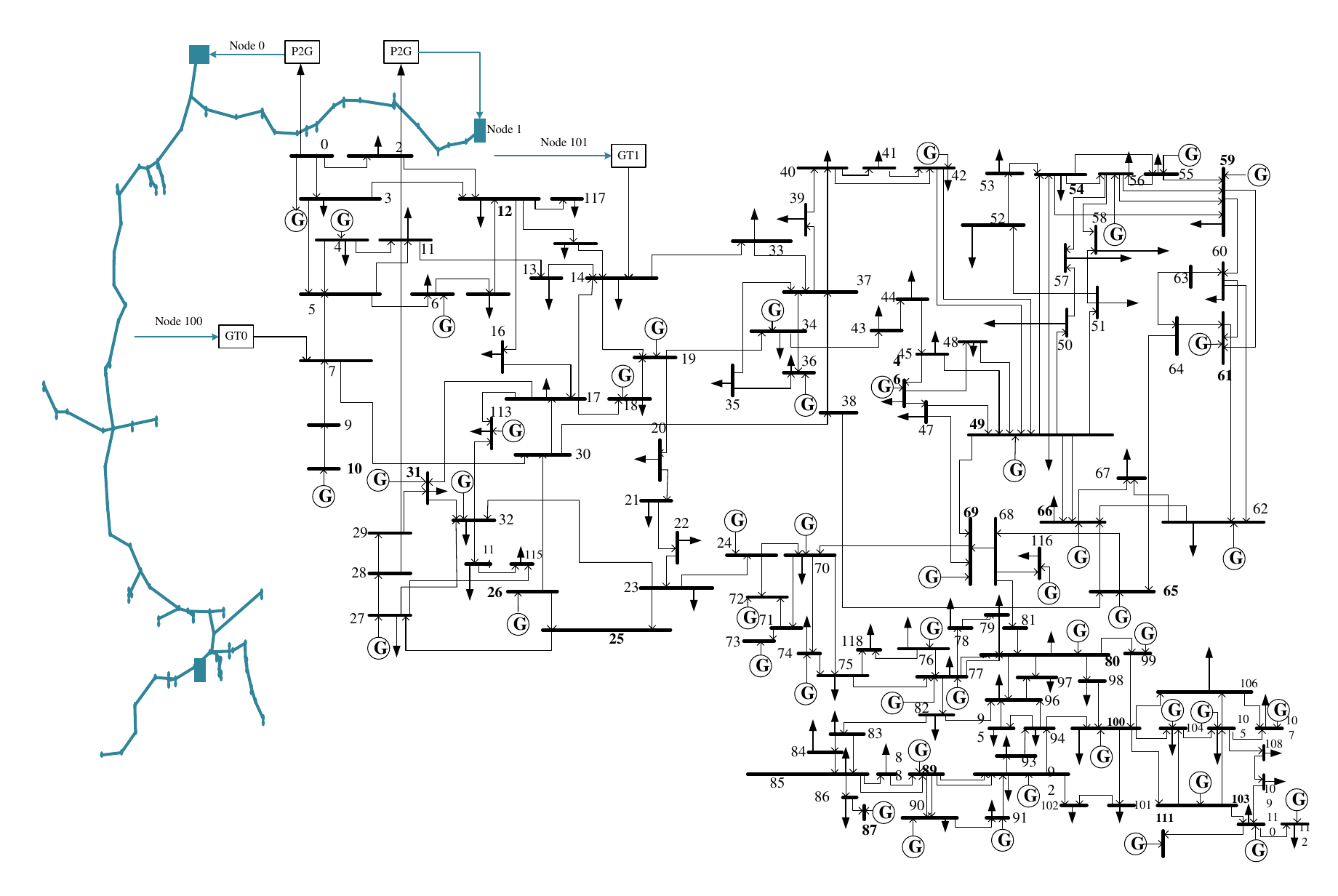}
    \caption{System diagram.}
    \label{case4-diagram}
\end{figure}

{The performance of the proposed method in large-system case was tested in the system shown in Fig. \mbox{\ref{case4-diagram}}. The system was composed of a 134-node NGS and a 118-bus EPS. The connections of the four coupling units can be found in TABLE \mbox{\ref{COUPLINGUNIT}}. Also, there were another three synchronous machines locating at bus 11, 39, and 68. The detailed parameters can be found in \mbox{\cite{caseparameter}}.}

{We set the leakage fault at the middle of pipe 58 with the leakage diameter $d=0.6D$ at $t=$300s and stopped the simulation when both GT0 and GT1 were tripped. The $\Delta x$ was set to be 100m. As a result, the system dimensions were over 80 thousand. We performed the previous five methods to locate the simulation events and profiled the time costs in TABLE \mbox{\ref{ELR3}} and \mbox{\ref{Time Costs3}}.}
\begin{table}[]
	\centering
        \caption{Coupling Units}
\begin{tabular}{lcc}
\hline
\multicolumn{1}{c}{} & Electric Bus & Gas Node \\ \hline
P2G0                 & 0            & 0        \\
P2G1                 & 28           & 1        \\
GT0                  & 7            & 100      \\
GT1                  & 14           & 101      \\ \hline
\end{tabular}
\label{COUPLINGUNIT}
\end{table}

{In this scenario, EU-TZ and CD-TZ methods failed due to divergence issues. The proposed WN-RD was able to locate the simulation events accurately when compared with the Benchmark, with a more than 18-fold speed up. While the KT-RD method was faster, it suffered from poor accuracy.}

\begin{table}[!h]
	\centering
        \caption{Event Location Results in Large System}
        \begin{tabular}{lcc}
        \hline
        Method& GT1 Tripping&GT0 Tripping \\ \hline
        Benchmark  &1691.18s  &2008.83s  \\
        EU-TZ   & Diverge &N\textbackslash A \\
        CD-TZ  & Diverge &N\textbackslash A\\
        KT-RD   & 1586.95s &1864.65s    \\
        WN-RD & 1684.78s &2008.87s  \\
        \hline
        \end{tabular}
        \label{ELR3}
\end{table}
\begin{table}[!h]
	\centering
        \caption{Time Costs in Large System}
        \begin{tabular}{lc}
	\hline
        Method& Total \\
        \hline
        Benchmark&23583.93s\\
        EU-TZ&N\textbackslash A\\
        CD-TZ&N\textbackslash A\\
        KT-RD&81.57s\\
	WN-RD&1251.09s\\ 
 \hline
	\end{tabular}
	\label{Time Costs3}
\end{table}
\section{Conclusion}
This paper uses simulation to study the cross-system propagation speed of the rupture and leakage faults in gas pipes. 
A semi-implicit simulation approach, together with a critical-time-location strategy, is proposed to perform efficient and robust post-fault co-simulation of the NGSs and EPSs.

In case studies, we found that {methods based on finite difference} are not good candidates for simulation because of the {oscillations, low-order, computational intractability, and non-convergence issues}. Instead, to perform spatial discretizations of PDEs by WENO-3 and solve the DAEs with the stiffly accurate Rosenbrock method can not only maintain the precision but also speed up the simulations.
Moreover, it was found that the intra-NGS fault propagation could last for hours. 
But the NGS faults within \SI{15}{\kilo\metre} range of EPS could affect EPS within \SI{10}{\minute} {and even trigger cascading failures of GTs}. The system operators should stay alert to critical facilities in this area.
The above-mentioned results {underline the importance of the post-fault co-simulation considering both gas and electric dynamics}. In addition, the results highlight the great potential of the proposed models and simulation approach to serve as a security analysis tool of the joint operation of NGSs and EPSs. 

{The major limitation of our study is that the proposed fault models cannot depict the dynamic evolution of post-fault systems in the event of pressure's zero-crossing. In this case, the load and pipe modeling should be extensively investigated by computation fluid dynamics and physical experiments, which will be our future research topics.}

\bibliographystyle{IEEEtran}
\bibliography{abrv,bibliography}
\end{document}